\documentclass[iop,numberedappendix]{emulateapj-rtx4}
\usepackage{graphicx,natbib,color,bm,url,times,url}
\graphicspath{{./fig/}{./png/}}

\newcommand{\EQ}{\begin{equation}}
\newcommand{\EN}{\end{equation}}
\newcommand{\EQA}{\begin{eqnarray}}
\newcommand{\ENA}{\end{eqnarray}}

\newcommand{\Eq}[1]{Equation~(\ref{#1})}
\newcommand{\Eqs}[2]{Equations~(\ref{#1}) and~(\ref{#2})}

\newcommand{\Sec}[1]{Section~\ref{#1}}

\newcommand{\Fig}[1]{Figure~\ref{#1}}

\newcommand{\Figp}[2]{Figure~\ref{#1}({#2})}

\newcommand{\Figs}[2]{Figures~\ref{#1} and \ref{#2}}

\newcommand{\Tab}[1]{Table~\ref{#1}}
\newcommand{\Tabs}[2]{Tables~\ref{#1} and \ref{#2}}

\newcommand{\bra}[1]{\langle #1\rangle}

{}
{}
{}

{}
{}
{}
{}
{}
{}
{}
{}
{}
{}
{}
{}
{}
{}
{}
{}
{}

{}

\newcommand{\meanB}{\overline{B}}

{}

{}
{}
{}

\newcommand{\dd}{{\rm d} {}}

\def\Teff{T_{\rm eff}}

\newcommand{\G}{\,{\rm G}}

\newcommand{\days}{\,{\rm d}}

\newcommand{\dex}{\,{\rm dex}}

\newcommand{\dday}{\,{\rm d}}
\newcommand{\yr}{\,{\rm yr}}

\newcommand{\Gyr}{\,{\rm Gyr}}

\newcommand{\yaj}[3]{ #1, {AJ,} {#2}, #3}

\newcommand{\yapj}[3]{ #1, {ApJ,} {#2}, #3}
\newcommand{\ypasp}[3]{ #1, {PASP,} {#2}, #3}
\newcommand{\yapjl}[3]{ #1, {ApJL,} {#2}, #3}

\newcommand{\yan}[3]{ #1, {Astron.\ Nachr.,} {#2}, #3}

\newcommand{\yana}[3]{ #1, {A\&A,} {#2}, #3}
\newcommand{\yanal}[3]{ #1, {A\&AL,} {#2}, #3}

\newcommand{\yass}[3]{ #1, {Ap\&SS,} {#2}, #3}

\newcommand{\yaraa}[3]{ #1, {ARA\&A,} {#2}, #3}

\newcommand{\ymn}[3]{ #1, {MNRAS,} {#2}, #3}
\newcommand{\ynat}[3]{ #1, {Nature,} {#2}, #3}
\newcommand{\ysci}[3]{ #1, {Science,} {#2}, #3}

\newcommand{\yproc}[5]{ #1, in {#3}, ed.\ #4 (#5), #2}

\newcommand{\sapj}[2]{ #1, {ApJ}, submitted, arXiv:#2}

\newcommand{\psph}[2]{ #1, {Sol.\ Phys.}, in press, arXiv:#2}

\begin{document}
\title{Evolution of coexisting long and short period stellar activity cycles}
\author{Axel Brandenburg$^{1,2,3,4}$, Savita Mathur$^5$, \& Travis S.\ Metcalfe$^5$
}
\affil{
$^1$Laboratory for Atmospheric and Space Physics, University of Colorado, Boulder, CO 80303, USA\\
$^2$JILA and Department of Astrophysical and Planetary Sciences, University of Colorado, Boulder, CO 80303, USA\\
$^3$Nordita, KTH Royal Institute of Technology and Stockholm University, Roslagstullsbacken 23, SE-10691 Stockholm, Sweden\\
$^4$Department of Astronomy, AlbaNova University Center, Stockholm University, SE-10691 Stockholm, Sweden\\
$^5$Space Science Institute, 4750 Walnut Street, Suite 205, Boulder, CO 80301, USA
}

%\submitted{\today, $ $Revision: 1.175 $ $}
\submitted{Astrophys. J. 845, 79 (2017)}
\date{Received 2017 April 28; revised 2017 June 27; accepted 2017 June 28; published 2017 August 14}

\begin{abstract}
The magnetic activity of the Sun becomes stronger and weaker over
roughly an 11 year cycle, modulating the radiation and charged particle
environment experienced by the Earth as ``space weather.'' Decades of
observations from the Mount Wilson Observatory have revealed that other stars
also show regular activity cycles in their Ca~{\sc ii}~H+K line
emission, and identified two different relationships between the length
of the cycle and the rotation rate of the star. Recent observations at
higher cadence have allowed the discovery of shorter cycles with periods
between $1$--$3\yr$. Some of these shorter cycles coexist with longer
cycle periods, suggesting that two underlying dynamos can operate
simultaneously. We combine these new observations with previous data,
and we show that the longer and shorter cycle periods agree remarkably
well with those expected from an earlier analysis based on the mean 
activity level and the rotation period. The relative turbulent length
scales associated with the two branches of cyclic behavior suggest that
a near-surface dynamo may be the dominant mechanism that drives cycles
in more active stars, whereas a dynamo operating in deeper layers may
dominate in less active stars.
However, several examples of equally prominent long and short cycles have
been found at all levels of activity of stars younger than $2.3\Gyr$.
Deviations from the expected cycle
periods show no dependence on the depth of the convection zone
or on the metallicity. 
For some stars that exhibit longer cycles, we
compute the periods of shorter cycles that might be detected with future
high-cadence observations.
\end{abstract}

\email{brandenb@nordita.org}

\section{Introduction}

Activity cycles akin to the 11 year sunspot cycle have been seen for
many late-type stars in their Ca~{\sc ii}~H+K line emission, since
\cite{Wil63,Wil68,Wil78} initiated their systematic study half a
century ago.
This emission is a proxy of chromospheric activity \citep{ES13}.
The Ca~{\sc ii}~H+K flux, normalized by the bolometric flux, is denoted
by $R_{\rm HK}$. The chromospheric contribution, denoted
by $R'_{\rm HK}$, is approximately proportional to the square root of
the mean magnetic field strength at the stellar surface \citep{Sch89}.
\cite{Noy83} showed that the time-averaged value, $\bra{R'_{\rm HK}}$,
is proportional to the inverse Rossby number, $\tau/P_{\rm rot}$,
where $\tau$ is the turnover time obtained from stellar mixing length
models and $P_{\rm rot}$ is the stellar rotation period.
For large values of $\tau/P_{\rm rot}$, the star's activity saturates at
$\log\bra{R'_{\rm HK}}\approx-4.2$ \citep{Noyes84}, which is interpreted
in terms of starspots having filled the entire surface of the star
\citep{SL85}.
The time trace of $R'_{\rm HK}$ varies with the activity cycle,
revealing its cycle period $P_{\rm cyc}$. However, it is affected by flares
and the presence of spots.
The relative cycle amplitude can be determined as
$A_{\rm cyc}=\Delta R'_{\rm HK}/ R'_{\rm HK}\equiv\Delta\ln R'_{\rm HK}$.
The time trace can also be used to infer $P_{\rm rot}$.
Understanding the dependence of these quantities on other stellar properties
is an important goal of stellar dynamo theory.

The work of \cite{NWV84} showed that, for G and K~dwarfs with $\tau$
between 11 and 26 days and $P_{\rm rot}$ between 22 and 48 days,
$1/P_{\rm cyc}$ scales with $\tau/P_{\rm rot}$ as
\EQ
1/P_{\rm cyc} \propto (\tau/P_{\rm rot})^n,
\label{scl_rel}
\EN
where $n=1.25$.
\cite{KRS83} theoretically studied two types of stellar dynamos:
one with overlapping induction layers \citep[as in the
early work of][]{Par55} and one with non-overlapping ones
\citep[as in the spherical shell models of][]{SK69}.
They found that, in the former case of overlapping induction layers,
the cycle period of the fastest growing mode obeys $n=4/3$,
which is close to the observed value.
Later, using nonlinear, plane-wave
dynamo models, \cite{RD82} found $n<1$ for different
nonlinearities, which was in conflict with the observations.
In a subsequent review, \cite{BV85} showed that there are many more stars
that do not obey any relation between $P_{\rm cyc}$ and $P_{\rm rot}$.
However, this view has changed considerably in subsequent years.

A major difficulty lies in the fact that, when comparing stars of
different spectral type, one has to rely on a meaningful determination
of $\tau$, which is a model-dependent quantity.
\cite{Noyes84} computed $\tau$ as the ratio of the mixing length
to the turbulent velocity approximately one scale height above the bottom
of the convection zone.
The relevance of such a definition is unclear, given that the location
of the dynamo is still not known.
\cite{Par75} assumed it to operate at the bottom of the convection
zone, whereas \cite{Bra05} argued in favor of a dynamo distributed throughout
the convection zone, but with the near-surface shear layer playing an important
role in producing the observed equatorward migration of the toroidal
flux belts; see corresponding models by \cite{PK11}.
Flux transport dynamos \citep{CSD95,DC99}, on the other hand,
are intermediate in the sense that the toroidal field resides mostly
at the bottom of the convection zone, and is the result of shearing
a poloidal field that is sourced by the tilt of decaying
active regions at the surface.
Thus, it has non-overlapping induction layers, as in \cite{SK69},
but with meridional circulation playing an important role in determining
cycle period and migration direction of the toroidal flux belts.
The resulting cycle period then decreases with increasing rotation
period with $n\approx-0.25$, which has the wrong sign \citep{JBB10,KKC14}.

To avoid the dependence on $\tau$ in the interpretation of stellar cycle
data, \cite{TRB88} considered the ratio $P_{\rm cyc}/P_{\rm rot}$, which
they found to decrease systematically with increasing thickness of the
convection zone.
Following a similar idea, \cite{SBZ93} plotted this ratio versus age and
color $B-V$, whereas \cite{BST98} (hereafter BST) plotted it versus both
$\bra{R'_{\rm HK}}$ (hereafter BST diagram) and $\tau/P_{\rm rot}$.
These papers used a subset of the extended data set from the Mount Wilson
HK project \citep{Bal+95}\footnote{\url{http://www.nso.edu/node/1335}}.
Owing to the proportionality between $\bra{R'_{\rm HK}}$ and
$\tau/P_{\rm rot}$, both graphs are similar and show two branches
with positive slopes, $\nu_{\rm A}$ and $\nu_{\rm I}$.
The subscripts A and I for active and inactive refer to the two groups
of stars in a previously identified bimodal distribution as a function
of $\bra{R'_{\rm HK}}$ with a local minimum at
$\log\bra{R'_{\rm HK}}\approx-4.75$,
which is known as the Vaughan--Preston gap \citep{VP80}.

The question of which quantities to plot against each other is of
considerable importance to the present paper.
The sensitivity to a model-dependent definition of the turnover time is
obviously alleviated when plotting just $P_{\rm cyc}$ against $P_{\rm rot}$,
as done by \cite{BV07} (hereafter BV), referred to below as the BV diagram.
In such a representation, the same two branches are recovered, suggesting a
basic equivalence between the BV and BST diagrams.
However, in the BV diagram, the Sun appears between the two branches,
whereas it is closer to the inactive branch in the BST diagram.
The idea that the Sun takes a special position (in one of the two
representations) has received much interest and deserves a renewed look.

Because $R'_{\rm HK}$ and $\tau/P_{\rm rot}$ are proportional to each other
and since $R'_{\rm HK}$ is often not available \citep{SB99}, the
representation of the period ratio over $\tau/P_{\rm rot}$ has become popular.
However, in comparison with the BV representation, it suffers from the
shortcoming that it now involves the model-dependent and ill-determined
turnover time.
It is therefore important to remind the reader that this criticism does not
apply to the original BST diagram, which does not use $\tau/P_{\rm rot}$,
but instead $R'_{\rm HK}$, which is an observational quantity -- independent
of stellar models.

An important difference between the BST and BV diagrams lies in the fact
that the BST diagram takes into account the dependence on another
parameter, namely $\bra{R'_{\rm HK}}$.
Using the linear representation of $P_{\rm cyc}$ versus $P_{\rm rot}$,
BV found that the two are proportional to each other, but with
different slopes.
In other words, their ratio $P_{\rm cyc}/P_{\rm rot}$ is constant on
each of the two branches.
If this were strictly true, the branches in the BST diagram should be
horizontal, so the slopes $\nu_{\rm I}$ and $\nu_{\rm A}$ would vanish.
This does not seem to be the case, however, and the new data points
discussed in this paper confirm this.
Indeed, a linear fit in the BV representation implies $n=1$ in
\Eq{scl_rel}, whereas \cite{NWV84} found $n=1.25$.

The graph of $P_{\rm rot}/P_{\rm cyc}$ versus $\bra{R'_{\rm HK}}$ has
a theoretical interpretation in terms of mean-field dynamo theory.
For this discussion, it is useful to define the cycle and rotation frequencies,
$\omega_{\rm cyc}=2\pi/P_{\rm cyc}$ and $\Omega=2\pi/P_{\rm rot}$,
so $\omega_{\rm cyc}/\Omega=P_{\rm rot}/P_{\rm cyc}$.
Next, we assume that both the $\alpha$ effect and the radial angular
velocity gradient,
$\Omega'$, are proportional to $\Omega$, and that the $\alpha$ effect
depends on the mean magnetic field, $\meanB$, via a ``quenching function.''
For a plane-wave $\alpha\Omega$ dynamo, the cycle
frequency is given by \citep{Sti76}
\EQ
\omega_{\rm cyc}\approx|\alpha\Omega'|^{1/2}.
\label{omcyc}
\EN
Because $R'_{\rm HK}$ is approximately proportional to $\meanB^{0.5}$
\citep{Sch89}, the graph of $\omega_{\rm cyc}/\Omega$ versus
$\bra{R'_{\rm HK}}$ is a direct representation of the quenching function.

Positive slopes, i.e., positive values of $\nu_{\rm A}$ and $\nu_{\rm I}$
therefore indicate that the $\alpha$ effect is an {\em increasing} function
of magnetic field strength on these branches.
This is referred to as ``anti-quenching,'' which suggests that the
$\alpha$ effect is magnetically driven, and therefore the result of a
magnetic instability \citep{BS98}---not, as usually assumed, driven
by flow instabilities such as convection, which would be suppressed
with increasing $\meanB$.
In this model, saturation of the dynamo would be achieved by turbulent
diffusion also being an increasing function of magnetic field strength (BST).
Explicit evidence for an $\alpha$ effect and a turbulent diffusivity
that increase with
increasing field strength has been obtained by \cite{CMRB11}.
For larger field strengths, however, we expect proper quenching
with a negative slope, which was confirmed for a group
of stars that was referred to as superactive stars \citep{SB99}.

The topic of stellar cycle frequencies has received increased attention
in recent years with new data of high cadence becoming available,
allowing the determination of shorter cycle periods 
\citep{Metcalfe10,Metcalfe13,Egeland15}.
Spectropolarimetric surveys also
led to measuring magnetic fields \citep{Marsden14}.
Furthermore, photometric data 
collected by missions such as {\it Kepler} provided additional activity cycle
detections and the measurement of activity levels
\citep[e.g.][]{Garcia10,Garcia14,Mathur14,Salabert16a}.
The question of cycle frequencies has become particularly interesting
in connection with the recent discovery of reduced magnetic
braking of stars whose activity falls below a certain threshold
\citep{vanSaders2016}.
The Sun happens to fall close to this threshold value
($\log\bra{R'_{\rm HK}}\approx-4.95$), suggesting it therefore to be
within a state of transition \citep{MEvS16}.
This idea has so far only been discussed in the framework of the BV diagram,
in which the Sun takes a prominent position.
One of our aims is therefore to clarify the apparent contradiction to the BST diagram.
For this purpose, we combine the results from a larger data set, and
assess carefully the reliability of all the stars in our sample.
Furthermore, we initially consider only K~dwarfs, because
they tend to have more clearly defined cycle periods.
We then study F and G~dwarfs---keeping in mind, however, that their cycles
tend to be more irregular, and therefore the determination of cycle
periods is often less certain.
We then discuss individual stars, first those with two cycles, and
then inactive and active stars with just one short or one long cycle,
respectively.
In several cases, we discuss theoretically computed values
that would be expected based on our fits.
Finally, we consider possible dependencies of the residuals on
metallicity, age, or thickness of the convection zone.

\newpage

\section{Analysis of cycle periods}

\subsection{Sample selection}

Out of the 112 stars of \cite{Bal+95}, BST used a subset of 21 stars.
BV excluded HD~219834A, but included 6 additional stars (HD~1835, HD~20630,
HD~76151, HD~100180, HD~190406, and HD~165341A), all of which were 
among the expanded sample of \cite{SB99}, who excluded HD~219834A because
the cycle period was comparable to the length of the data set.
Except for the Sun, the remaining 20 stars of BST were all part
of the 50 stars in the original program of \cite{Wil78}.
Plots of the time series and associated short-term Fourier transforms
of many of these stars can be found in \cite{Olah_etal16}.
In the following, we analyze the cycle periods of 35 stars,
which include the 27 stars from \cite{Bal+95},
two additional stars from their sample (HD~22049 and HD~30495), and
two stars observed by the {\it Kepler} mission---KIC~10644253 and KIC~8006161
\citep{Salabert16a,Kiefer17}. We also include HD~128620 and HD~128621
\citep[=~$\alpha$~Cen A \& B,][]{Ayres14},
along with the solar analogs HD~146233 \citep[=~18~Sco;][]{Hall07b}
and HD~17051 \citep[$\iota$~Horologii;][]{Metcalfe10}.

The sample of \cite{SB99} contains another 30 active and
inactive stars from \cite{Bal+95} and one more photometric variable,
as well as 28 superactive stars with periods up to $60\yr$.
\cite{Lehtinen16} discovered secondary periods in some superactive stars
in his sample of 21 additional stars.
In the following, however, we restrict ourselves to the reduced
sample of active and inactive stars, including those with short
periods in the $1$--$3\yr$ range.
It should be noted, however, that
for a few stars from the {\it CoRoT} mission, \cite{FerreiraLopes15}
claim to have found cycles even shorter than one year, but the data
record is too sparse to be reliable.
Their detailed study is beyond the scope of the present paper and
requires a larger sample, including stars from the {\it Kepler} mission;
see \cite{Reinhold} and \cite{Montet} for steps in that direction.

\subsection{Representation of and fits to the data}

Following BST, we plot $\omega_{\rm cyc}/\Omega$ versus $\bra{R'_{\rm HK}}$
and determine the parameters to the fit
\EQ
\omega_{\rm cyc}/\Omega = b_i \bra{R'_{\rm HK}}^{\nu_i}.
\label{oO_RHK_relation}
\EN
BST found $\nu_{\rm I}=0.85$ for inactive stars and
$\nu_{\rm A}=0.72$ for active stars.
In \Eq{oO_RHK_relation}, the index $i$ stands for I (for inactive)
or A (for active).
Occasionally, we refer to these fits as short and long period branches,
which is more compatible with the notion that inactive and active stars
can have both cycle periods coexisting.
Below, however, short cycle periods are found to range from
$1.5\yr$ to $21\yr$, whereas long ones start already at $5.6\yr$,
so the attributes long and short are only relative.

In the case of the Sun, which will be considered in \Sec{Gdwarfs} along
with other G~dwarfs, the longer period might correspond to the period
of the Gleissberg cycle,
which is about seven times longer than the basic 11 year solar cycle.
This is too long to be determined for other stars with the currently
available data sets.
However, there are several stars whose basic cycle period
is in the $1$--$3\yr$ range.
Many of them also have a longer period in the $10$--$20\yr$ range.
These fit well onto the two lines in the BST diagram.

For each of the two branches, the values of $b_i$ are non-intuitive,
because they correspond to the value of $\omega_{\rm cyc}/\Omega$ for
$\log R'_{\rm HK}=0$, which is far away from realistic values.
Therefore, it is not meaningful to compare $b_i$, because it
varies significantly even for small changes in $\nu_i$.
Thus, we compute
\EQ
\tilde{b}_i=b_i R_{\rm HKvp}'^{\;\nu_i}\;,
\label{btilde}
\EN
which corresponds to the value of $\omega_{\rm cyc}/\Omega$ at the
Vaughan--Preston gap at $\log R'_{\rm HK}=\log R_{\rm HKvp}'=-4.75$.
These values are listed in \Tab{fitcoefs} along with $\nu_i$.

BST also consider the dependence of $\omega_{\rm cyc}/\Omega$ on
$\tau/P_{\rm rot}$, which is similar to that on $\bra{R'_{\rm HK}}$,
because $\bra{R'_{\rm HK}}$ is well correlated with $\tau/P_{\rm rot}$
\citep{Noy83}.
Within the range of interest for cyclic dynamos, i.e., for
$\log\bra{R'_{\rm HK}}<-4.4$, BST found
\EQ
\bra{R'_{\rm HK}} = c\, (\tau/P_{\rm rot})^\mu,
\label{RHK_relation}
\EN
with $\mu=0.99$, independent of which of the two branches
the cycle frequencies lie on; see Figure~1(c) of BST.
Because $\mu$ is close to unity, we can compute the value of $c$ simply
from $\log c=\bra{\log\bra{R'_{\rm HK}}-\log(\tau/P_{\rm rot})}_{\rm sample}
\approx-4.631$ for the entire sample of stars.

Based on the possible coexistence of two branches for a given value of
$\bra{R'_{\rm HK}}$, we can compute values of $\omega_{\rm cyc}/\Omega$
that would fall perfectly on either of the two branches.
We refer to these as ``computed'' cycle periods.
Departures from those values could be
explained by a dependence on additional parameters, such as the
depth of the convection zone $d$ normalized by the stellar radius $R$, 
the metallicity [Fe/H], or the age, which might be affecting the cycle period.

In the present paper, we also compare with the BV diagram.
When comparing trends in such diagrams, it is useful
to convert the fits in the BST diagram using
\Eqs{oO_RHK_relation}{RHK_relation}, i.e.,
\EQ
P_{\rm rot}/P_{\rm cyc}\equiv\omega_{\rm cyc}/\Omega
=b_i\,\bra{R'_{\rm HK}}^{\nu_i}=C_i/P_{\rm rot}^{\mu\nu_i},
\label{conversion_formula}
\EN
where we have defined $C_i=b_i (c\tau^\mu)^{\nu_i}$
and $\tau(B-V)$ is taken from \cite{Noyes84}, i.e.,
\EQ
\log\tau=\left\{
\begin{array}{l}
1.362-0.166x+0.03x^2-5.3x^3,\;\; x>0,\;\\
1.362-0.14x,\quad x<0,
\end{array}
\right.
\EN
with $x=1-(B-V)$.
Thus, we have
\EQ
P_{\rm cyc}=P_{\rm rot}^{1+{n_i}}/C_i,
\label{conversion}
\EN
which will be overplotted below in our BV diagrams for $i={\rm A}$
and ${\rm I}$, as well as given values of $\tau$.
In \Tab{fitcoefs} we give the values of $\nu_i$ and $\tilde{b}_i$ from BST,
as well as for our sample of K~dwarfs discussed in \Sec{Kdwarfs} and
those obtained when including our sample of F and G~dwarfs discussed
in \Sec{Gdwarfs}.

\begin{table}[b!]\caption{
Fit coefficients for \Eqs{oO_RHK_relation}{btilde}.
}\vspace{12pt}\centerline{\begin{tabular}{lcccc}
& BST & K~dwarfs & F,G~dwarfs & F,G,K~dwarfs \\
\hline
$\nu_{\rm I}$           &~~~~0.85 &~~~~0.75 &~~~~0.68 &~~~~0.45 \\
$\nu_{\rm A}$           &~~~~0.72 &~~~~0.95 &0.25,$\;$ 0.32 &~~~~0.44 \\
$\log\tilde{b}_{\rm I}$ & $-1.83$ & $-1.85$ & $-1.87$ & $-1.91$ \\
$\log\tilde{b}_{\rm A}$ & $-2.67$ & $-2.65$ &$-2.67,\;-2.44$ & $-2.63$ \\
\label{fitcoefs}\end{tabular}}\tablenotemark{
The two values of $\nu_{\rm A}$ and $\log\tilde{b}_{\rm A}$ for F,G~dwarfs
apply to the Aa and Ab branches of BV; see \Sec{Gdwarfs}.
}\end{table}

There are several outliers that have been ignored in the calculation
of the fit coefficients given in \Tab{fitcoefs}.
For HD~78366 (blue $g$ symbol) and HD~114710 (blue $j$ symbol),
the shorter of the two cycle periods still fall close to the A branch.
BV argued that their positions in the plot constitute a third branch,
which she called the Ab branch.
The situation is similar with the $19.2\yr$ period of HD~128620
(blue $k$ symbol) and the the $12.9\yr$ period of HD~100180
(blue $H$ symbol), both of which would fall on the Ab branch
and will be discussed separately in \Sec{Gdwarfs}.
Therefore, they are not used in the determination of the I or A branches.
We return to a detailed discussion of those stars in \Sec{Discussion},
where we argue that the shorter cycle periods close to the A~branch
are questionable.
We also exclude the estimated $80\yr$ period of the solar Gleissberg cycle
(blue $A$ symbol) from the fit to the A~branch.
We emphasize, however, that none of these stars have been excluded from
any of the plots.

\subsection{Stellar ages}
\label{StellarAges}

Knowing the star's age is important when interpreting stellar cycles
and especially the simultaneous appearance of two cycle periods.
There has been significant progress in determining ages
through gyrochronology and asteroseismology.
For most of the stars, we have computed their ages from Equations~(12)--(14)
of \cite{MH08} as
\EQ
t = \left\{P_{\rm rot}/ [0.407\,(B-V-0.495)^{0.325}]\right\}^{1.767}.
\EN
The age for HD~219834A is assumed to be the same as its cooler companion 
HD~219834B. We adopted asteroseismic ages for HD~17051 \citep{Vauclair2008}, 
HD~128620, HD~128621 \citep{Bazot2012}, HD~146233 \citep{Li2012,Mittag2016},
KIC~8006161, KIC~10644253 \citep{Creevey2017}, HD~186408 and HD~186427 
\citep{MCD15}.

\subsection{K~dwarfs}
\label{Kdwarfs}

In \Fig{pK}, we plot the frequency ratio, $\omega_{\rm cyc}/\Omega$, versus
$\bra{R'_{\rm HK}}$ for our sample of K~dwarfs with well-defined cycles.
We overplot the fits given by \Eq{oO_RHK_relation} using the coefficients
for all F, G, K~stars listed in the last column of \Tab{fitcoefs}.
The corresponding cycle periods are denoted by $P_{\rm cyc}^{\rm I}$
and $P_{\rm cyc}^{\rm A}$, depending on which of the two branches they
are closest to.
The stars are indicated correspondingly by lower- and uppercase symbols.

Most of the K~dwarfs seem to form a tight relation around lines that have
a somewhat larger slope than to the fits for the full set of F, G, K~dwarfs.
To quantify this further, we determine a separate fit for K~dwarfs, but
exclude here HD~22049 (red E/e symbols) on both branches, as well as
HD~165341A (red n symbol), HD~149661 (red k symbol), and
HD~219834A (red r symbol) on the I~branch.
The fit parameters are listed in \Tab{fitcoefs} under K~stars,
and correspond to lines with larger slopes than for the full set
of stars, where the aforementioned stars are {\em not} ignored.

Three of our K~dwarfs have two cycle periods.
The occurrence of two cycle periods is not uncommon and has been reported
in a number of earlier papers \citep[see, e.g.,][]{SB92,SBZ93}.
When possible, both periods are listed in \Tab{Kstars} together with
other properties of the K~dwarfs of our sample.
We use the relative cycle amplitudes, $A_{\rm cyc}$, obtained by \cite{SB02},
to indicate the range of variation in $R'_{\rm HK}$ during the cycle.
For $\alpha$ Cen A and B, we compared their X-ray cycles with that
of the Sun \citep{Ayres14} and estimated their cycle amplitudes to be
approximately three and two times smaller than for the Sun.

The spectroscopic parameters $T_{\rm eff}$ and [Fe/H] in \Tab{Kstars} are
mostly from the Geneva-Copenhagen Survey \citep{Nor+04} and \cite{Ramirez13}.
However, we obtained values from other sources for HD 201091 and 201092
\citep{Kervella08}, HD 219834A \citep{Gray+06},
and HD 219834B \citep{Fuhrmann08}.
For most stars, the activity
indexes and cycle periods are from \cite{Bal+95}.
Rotation periods and uncertainties come from \cite{Bal+96} and
\cite{Donahue96}.
Exceptions are discussed in \Sec{Discussion}.

It is instructive to see where the stars shown in \Fig{pK} are located
in the BV diagram.
The result is shown in \Fig{pKBV}.
The solid lines show the fits found by BV (hereafter BV lines), whereas
the curves correspond to the fit of BST, 
which has been converted into a $P_{\rm cyc}$ versus
$P_{\rm rot}$ dependence using \Eq{conversion} for three values of $\tau$.

Interestingly, in the BV plot, the two cycle periods of HD~22049
(red E/e symbols) do fall onto the two branches.
The same is true of HD~165341A (red N/n symbols), where the longer cycle period
of $15.5\yr$ agrees well with the computed one (see \Tab{Kstars}),
but the shorter one of $5.1\yr$ is twice as long as the computed one
of $2.8\yr$.
In the BV plot, on the other hand, this star falls onto the lower
(inactive) branch.

The two branches of BST correspond to a bundle of lines in the BV plot,
because the conversion from one to the other requires an assumption about
the value of $\tau$; see \Eq{conversion_formula} and the definition of
$C_i$, which involves $\tau$.
In the BV plot, the BST lines are not straight, but they would become
straight in a double-logarithmic version of this plot.
Some of the scatter can be explained by the fact that the K~dwarfs have
different values of $\tau$.

\begin{table*}[t]\caption{
Sample of K~dwarfs (red symbols).
}\vspace{12pt}\centerline{\begin{tabular}{cccccrcccrccrcrcc}
$\!\!\!$Sym$\!\!\!$& HD/KIC & Sp & $B$--$V$ & $\Teff$ & [Fe/H]$\!$ & $d/R$ & $\!\!\log\bra{R'_{\rm HK}}\!\!$ &
age & $\tau\;$ & $P_{\rm rot}$ & $P_{\rm cyc}^{\rm I}$ & comp &
$P_{\rm cyc}^{\rm A}$ & comp &
$A_{\rm cyc}^{\rm I}$ & $A_{\rm cyc}^{\rm A}$ \\
\hline
% :.,.+18s/    /\~\~\~\~\~/g
% :.,.+18s/   / \~\~/g
% :.,.+18s/  1653411/\~\~\~165341A/g
% :.,.+18s/  2198341/\~\~\~219834A/g
% :.,.+18s/  2198342/\~\~\~219834B/g
% :.,.+18s/\\pm\*\*\*/      /g
% :.,.+18s/\*\*\*\*/    /g
% label      HD  Sp  BV  Teff   Fe/H    d/R   RHK'  age tau Prot Pcyc1 comp Pcyc2 comp Acyc Acyc2
{\rm a}&~~~~~3651&K0&0.84&5128&$ 0.19$&0.327&$-4.99$&7.2&20.6&$44.0      $ &$13.8\pm0.4$&12.6 &$          $&66.5&0.36&    \\
{\rm b}&~~~~~4628&K2&0.89&5035&$-0.17$&0.303&$-4.85$&5.3&21.7&$38.5\pm2.1$ &$ 8.6\pm0.1$& 9.6 &$          $&50.5&0.38&    \\
{\rm c}& ~~10476&K1&0.84&5188&$-0.04$&0.317&$-4.91$&4.9&20.6&$35.2\pm1.6$ &$ 9.6\pm0.1$& 9.3 &$          $&49.1&0.38&    \\
{\rm d}& ~~16160&K3&0.98&4819&$ 0.08$&0.326&$-4.96$&6.9&22.8&$48.0\pm4.7$ &$13.2\pm0.2$&13.3 &$          $&70.1&0.32&    \\
{\rm e}& ~~22049&K2&0.88&5152&$ 0.00$&0.319&$-4.46$&0.6&21.5&$11.1\pm0.1$ &$ 2.9\pm0.1$& 1.8 &$12.7\pm0.3$& 9.7&    &    \\
{\rm f}& ~~26965&K1&0.82&5284&$-0.04$&0.314&$-4.87$&7.2&20.1&$43.0      $ &$10.1\pm0.1$&10.9 &$          $&57.6&0.38&    \\
{\rm g}& ~~32147&K5&1.06&4745&$ 0.19$&0.354&$-4.95$&6.3&23.5&$48.0      $ &$11.1\pm0.2$&13.2 &$          $&69.4&0.42&    \\
{\rm h}& ~~81809&K0&0.80&5623&$-0.29$&0.305&$-4.92$&6.6&19.4&$40.2\pm3.0$ &$ 8.2\pm0.1$&10.7 &$          $&56.6&    &    \\
{\rm i}&  115404&K1&0.93&5081&$-0.16$&0.306&$-4.48$&1.4&22.3&$18.5\pm1.3$ &$          $& 3.1 &$12.4\pm0.4$&16.6&    &0.16\\
{\rm j}&  128621&K1&0.88&5230&$ 0.27$&0.339&$-4.93$&5.4&21.5&$36.2\pm1.4$ &$ 8.1\pm0.2$& 9.7 &$          $&51.4&0.11&    \\
{\rm k}&  149661&K2&0.80&5199&$-0.01$&0.302&$-4.58$&2.1&19.4&$21.1\pm1.4$ &$ 4.0\pm0.1$& 4.0 &$17.4\pm0.7$&21.0&0.35&0.15\\
{\rm l}&  156026&K5&1.16&4600&$-0.34$&0.311&$-4.66$&1.3&24.2&$21.0      $ &$          $& 4.3 &$21.0\pm0.9$&22.7&    &0.37\\
{\rm m}&  160346&K3&0.96&4797&$-0.09$&0.335&$-4.79$&4.4&22.7&$36.4\pm1.2$ &$ 7.0\pm0.1$& 8.5 &$          $&45.0&0.44&    \\
{\rm n}&~~~165341A&K1&0.78&5023&$-0.29$&0.307&$-4.55$&2.0&18.6&$19.9      $ &$ 5.1\pm0.1$& 3.6 &$15.5    $&19.1&0.54&0.12\\
{\rm o}&  166620&K5&0.90&5000&$-0.08$&0.333&$-4.96$&6.2&21.9&$42.4\pm3.7$ &$15.8\pm0.3$&11.7 &$          $&61.8&0.30&    \\
{\rm p}&  201091&K5&1.18&4400&$-0.20$&0.338&$-4.76$&3.3&24.4&$35.4\pm9.2$ &$ 7.3\pm0.1$& 8.0 &$          $&42.4&0.32&    \\
{\rm q}&  201092&K7&1.37&4040&$-0.27$&0.369&$-4.89$&3.2&25.9&$37.8\pm7.4$ &$11.7\pm0.4$& 9.8 &$          $&51.6&0.21&    \\
{\rm r}&~~~219834A&K5&0.80&5461&$ 0.23$&0.321&$-5.07$&6.2&19.4&$42.0      $ &$21.0\pm1.0$&13.0 &$          $&68.5&    &    \\
{\rm s}&~~~219834B&K2&0.91&5136&$ 0.24$&0.342&$-4.94$&6.2&22.1&$43.0      $ &$10.0\pm0.2$&11.7 &$          $&61.9&0.29&    \\
\label{Kstars}\end{tabular}}
\tablenotemark{Note: $\Teff$ is in Kelvin, [Fe/H] is in dex,
age is in Gyr, $\tau$ and $P_{\rm rot}$ are in days, whereas
$P_{\rm cyc}^{\rm I}$ and $P_{\rm cyc}^{\rm A}$
along with their computed values are in years.}
\end{table*}

\subsection{F and G~dwarfs}
\label{Gdwarfs}

We now discuss F and G~dwarfs listed in \Tab{Gstars}.
Here again, most of the spectroscopic inputs come from the 
Geneva-Copenhagen Survey \citep{Nor+04}, whereas the rotation and activity
measurements and uncertainties are from \cite{Bal+95,Bal+96} and \cite{Donahue96}.
For HD~128620, the data come from \cite{Ramirez13}, \cite{Ayres14}, and
\cite{Bazot2007}. 
For the {\it Kepler} stars, the spectroscopic parameters were obtained by
\cite{BL15}, and the rotation periods come from \citet{Garcia14}.
For KIC~8006161, the activity index and cycle period were measured by Karoff 
et al.\ (submitted). For KIC~10644253, the magnetic activity index and cycle 
measurement were done by \citet{Salabert16a}.

In \Fig{pG} we plot $\omega_{\rm cyc}/\Omega$ versus $\log\bra{R'_{\rm HK}}$
for our sample of F and G stars.
We compare these data points with the fits given by \Eq{oO_RHK_relation}
using the coefficients listed in \Tab{fitcoefs}.
We see that there are now somewhat stronger departures than for the K~dwarfs,
particularly for low $\bra{R'_{\rm HK}}$ values.
On the other hand, the fraction of active F and G~dwarfs is larger
than for the sample of K~dwarfs.
The solar $11\yr$ cycle (blue $a$ symbol) now departs much more
from the fit than in the original diagram of BST, where the
shorter cycle period of $10\yr$ and a longer rotation period of
$26.1\days$ \citep{Bal+95} were used.
HD~128620 (blue $k$ symbol) is well below the fit line, but it is
also old ($5.4\Gyr$) and extremely inactive compared to all the other
inactive stars.
It will therefore be interesting to find out whether this departure
may be connected with the recent discovery of reduced magnetic
braking for sufficiently inactive stars \citep{vanSaders2016,MEvS16,MvS17},
which has been associated with the dynamo having become subcritical once
the rotation rate drops below a critical rotation rate \citep{KN17}.

\begin{figure}[t!]\begin{center}
\includegraphics[width=\columnwidth]{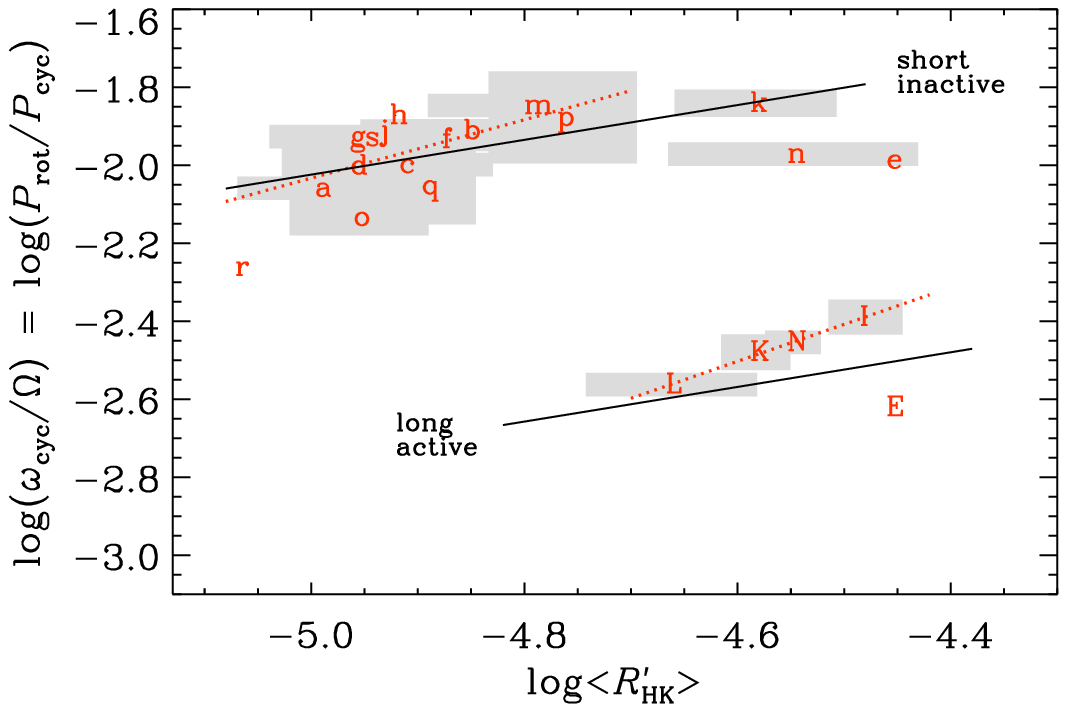}
\end{center}\caption[]{
$\omega_{\rm cyc}^{\rm I}/\Omega$ (lowercase symbols) and
$\omega_{\rm cyc}^{\rm A}/\Omega$ (uppercase symbols) versus
$\log\bra{R'_{\rm HK}}$ for the K~dwarfs listed in \Tab{Kstars}.
The upper and lower black solid lines come from fits
to all F, G, K~dwarfs, whereas red dashed lines
correspond to fits for K stars only (see text).
The gray boxes behind the letters indicate the
statistical error in the period ratio and the spread of
$\log R'_{\rm HK}$ between cycle minimum and maximum.
\vspace{12pt}}\label{pK}\end{figure}

\begin{figure}[t!]\begin{center}
\includegraphics[width=\columnwidth]{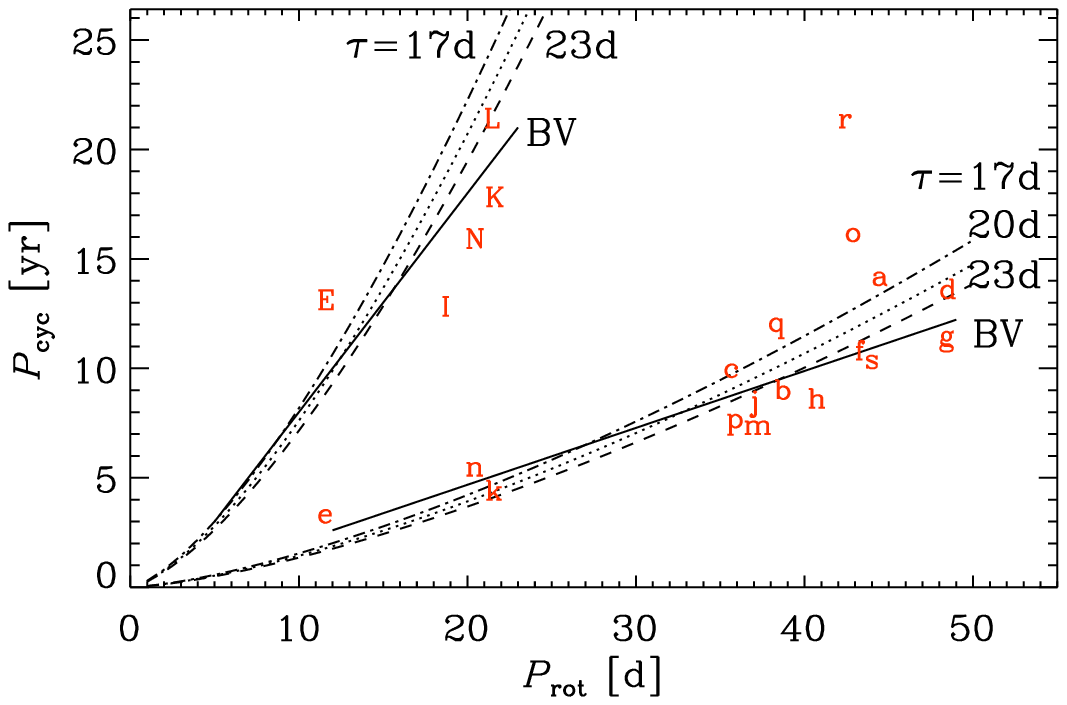}
\end{center}\caption[]{
$P_{\rm cyc}^{\rm I}$ (lowercase symbols) and $P_{\rm cyc}^{\rm A}$
(uppercase symbols) versus $P_{\rm rot}$ for the K~dwarfs listed in
\Tab{Kstars}.
The dash-dotted, dotted, and dashed lines correspond to $\tau=17\dday$,
$20\dday$, and $23\dday$, respectively.
\vspace{12pt}}\label{pKBV}\end{figure}

The BV plot for our sample of F and G~dwarfs is shown in \Fig{pGBV}.
It has more scatter than the plot for the K~dwarfs in \Fig{pKBV}.
However, systematic departures in the BST plot tend to correspond
to similar departures in the BV plot.
An example is HD~78366, where both the $g$ and $G$ symbols in the BST diagram
of \Fig{pG} are also close to the line for active stars.
This is similar for HD~114710, where the $j$ and $J$ symbols
are close to the line for active stars.
The blue $g$ and $j$ symbols fall along the Ab branch identified by BV.
However, the significance of this branch remains uncertain, as will be
discussed in \Sec{StarsWithTwoCyclePeriods}.
Our fit to this branch shown in \Fig{pG} (with parameters given in
\Tab{fitcoefs} under F,G stars) also includes the longer cycle periods
of HD~20630 (blue $D$ symbol) and HD~100180 (blue $H$ symbol) -- in
agreement with BV.
Interestingly, HD~128620 (blue $k$ symbol) also fits well on this line.
The corresponding Aa branch of BV includes the long cycle periods of
HD~114710 (blue $J$ symbol) and HD~78366 (blue $G$ symbol), as well as
those of HD~190406 (blue $N$ symbol), HD~30495 (blue $E$ symbol),
HD~152391 (blue $M$ symbol), and HD~1835 (blue $B$ symbol).
Next, we combine the data points of the Aa and Ab
branches with those of the long cycle periods of K~dwarfs, which lie
in between the Aa and Ab branches; see the coefficients
in \Tab{fitcoefs} under F,G,K~dwarfs, 
This was shown as the lower solid lines of \Figs{pK}{pG}.
The combined BST diagram is shown in \Fig{pKG} and will be
discussed in detail in \Sec{Discussion}.

\begin{table*}[t]\caption{
Sample of F and G~dwarfs (blue symbols).
}\vspace{12pt}\centerline{\begin{tabular}{crcccrcccrccrcrcc}
$\!\!\!$Sym$\!\!\!$& HD/KIC & Sp & $B$--$V$ & $\Teff$ & [Fe/H]$\!$ & $d/R$ & $\!\!\log\bra{R'_{\rm HK}}\!\!$ &
age & $\tau\;$ & $P_{\rm rot}$ & $P_{\rm cyc}^{\rm I}$ & comp &
$P_{\rm cyc}^{\rm A}$ & comp &
$A_{\rm cyc}^{\rm I}$ & $A_{\rm cyc}^{\rm A}$ \\
\hline
% :.,.+16s/       0/     Sun/g
% :.,.+18s/\\pm\*\*\*/      /g
% :.,.+18s/\*\*\*\*/    /g
% label      HD  Sp  BV  Teff   Fe/H    d/R   RHK'  age tau Prot Pcyc1 comp Pcyc2 comp Acyc Acyc2
{\rm a}&     Sun&G2&0.66&5778&$ 0.00$&0.292&$-4.90$&4.6&12.6&$25.4\pm1.0$ &$11.0\pm2.0$& 6.6 &$80.0      $&35.0&0.22&    \\
{\rm b}&    1835&G3&0.66&5688&$-0.02$&0.275&$-4.43$&0.5&12.6&$ 7.8\pm0.6$ &$          $& 1.3 &$ 9.1\pm0.3$& 6.6&    &0.15\\
{\rm c}&   17051&F8&0.57&6053&$ 0.00$&0.267&$-4.60$&0.6& 7.5&$ 8.5\pm0.1$ &$ 1.6      $& 1.6 &$          $& 8.6&    &    \\
{\rm d}&   20630&G5&0.66&5701&$ 0.00$&0.292&$-4.42$&0.7&12.6&$ 9.2\pm0.3$ &$          $& 1.5 &$ 5.6\pm0.1$& 7.8&    &0.14\\
{\rm e}&   30495&G1&0.63&5780&$-0.08$&0.263&$-4.49$&1.1&10.9&$11.4\pm0.2$ &$ 1.7\pm0.3$& 2.0 &$12.2\pm3.0$&10.3&    &    \\
{\rm f}&   76151&G3&0.67&5675&$-0.04$&0.301&$-4.66$&1.6&13.2&$15.0      $ &$ 2.5\pm0.1$& 3.1 &$          $&16.1&0.16&    \\
{\rm g}&   78366&G0&0.63&5915&$-0.10$&0.322&$-4.61$&0.8&10.9&$ 9.7\pm0.6$ &$ 5.9\pm0.1$& 1.9 &$12.2\pm0.4$& 9.9&0.19&0.27\\
{\rm h}&  100180&F7&0.57&5942&$-0.15$&0.221&$-4.92$&2.3& 7.5&$14.0      $ &$ 3.6\pm0.1$& 3.7 &$12.9\pm0.5$&19.7&0.07&0.17\\
{\rm i}&  103095&G8&0.75&5035&$-1.36$&0.096&$-4.90$&4.6&17.4&$31.0      $ &$ 7.3\pm0.1$& 8.1 &$          $&42.5&0.27&    \\
{\rm j}&  114710&F9&0.58&5970&$-0.06$&0.278&$-4.75$&1.7& 8.0&$12.3\pm1.1$ &$ 9.6\pm0.3$& 2.8 &$16.6\pm0.6$&14.5&0.12&0.21\\
{\rm k}&  128620&G2&0.71&5809&$ 0.23$&0.320&$-5.00$&5.4&15.4&$22.5\pm5.9$ &$19.2\pm0.7$& 6.5 &$          $&34.3&0.07&    \\
{\rm l}&  146233&G5&0.65&5767&$-0.02$&0.268&$-4.93$&4.1&12.0&$22.7\pm0.5$ &$ 7.1      $& 6.1 &$          $&32.2&    &    \\
{\rm m}&  152391&G7&0.76&5420&$-0.08$&0.325&$-4.45$&0.8&17.8&$11.4\pm1.4$ &$          $& 1.9 &$10.9\pm0.2$& 9.9&    &0.28\\
{\rm n}&  190406&G1&0.61&5847&$-0.12$&0.241&$-4.80$&1.8& 9.7&$13.9\pm1.5$ &$ 2.6\pm0.1$& 3.3 &$16.9\pm0.8$&17.3&0.15&0.34\\
{\rm o}& 8006161&G8&0.84&5488&$ 0.34$&0.359&$-5.00$&4.6&20.6&$29.8\pm3.1$ &$ 7.4\pm1.2$& 8.6 &$          $&45.3&    &    \\
{\rm p}&10644253&G0&0.59&6045&$ 0.06$&0.274&$-4.69$&0.9& 8.6&$10.9\pm0.9$ &$ 1.5\pm0.1$& 2.3 &$          $&12.1&    &    \\
{\em q}&  186408&G2&0.64&5741&$ 0.05$&0.278&$-5.10$&7.0&11.5&$23.8\pm1.7$ &$          $& 7.6 &$          $&40.2&    &    \\
{\em r}&  186427&G3&0.66&5701&$ 0.05$&0.291&$-5.08$&7.0&12.6&$23.2\pm3.2$ &$          $& 7.3 &$          $&38.4&    &    \\
\label{Gstars}\end{tabular}}
\tablenotemark{Note: $\Teff$ is in Kelvin, [Fe/H] is in dex,
age is in Gyr, $\tau$ and $P_{\rm rot}$ are in days, whereas
$P_{\rm cyc}^{\rm I}$ and $P_{\rm cyc}^{\rm A}$
along with their computed values are in years.}
\end{table*}

\begin{figure}[t!]\begin{center}
\includegraphics[width=\columnwidth]{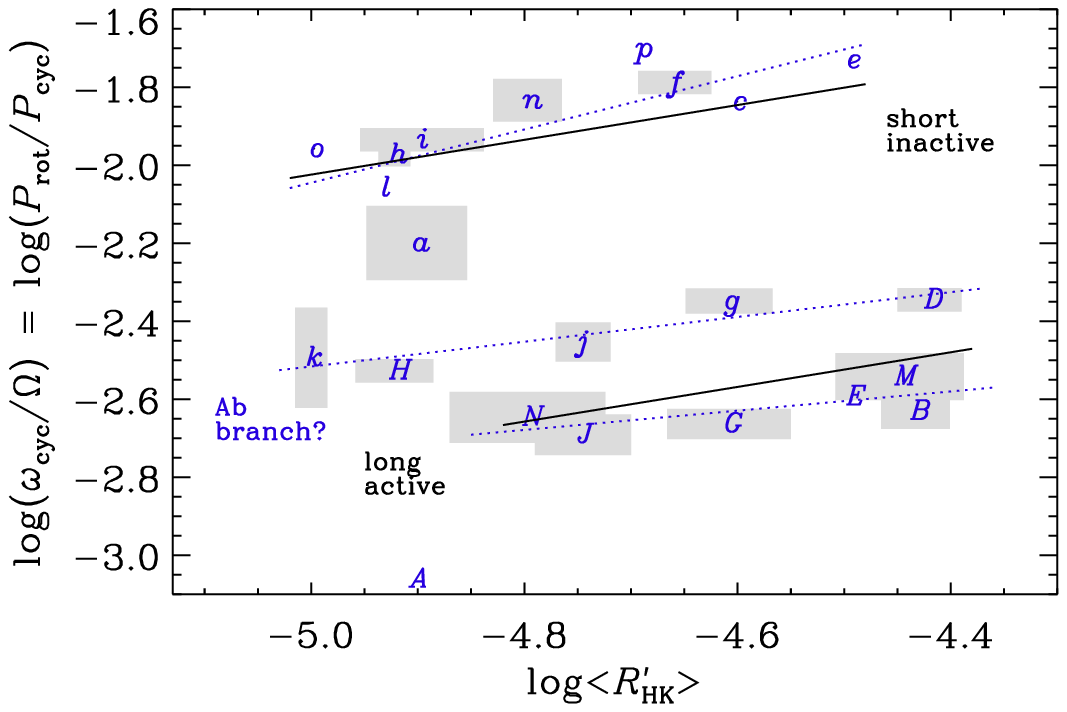}
\end{center}\caption[]{
BST diagram for the F and G~dwarfs listed in \Tab{Gstars}.
The black solid lines refer to our fits to F, G, and K stars
discussed in the text (and also shown in \Fig{pK})
whereas the upper blue dotted line denotes the fit to the
F and G stars. The two lower blue dotted lines denote what
corresponds to the Aa and Ab branches in BV, which are here
evaluated as fits through N, J, G, E, M, and B on the
one hand, and j and g together with D and H, on the other.
Otherwise like \Fig{pK}.
}\label{pG}\end{figure}

\begin{figure}[t!]\begin{center}
\includegraphics[width=\columnwidth]{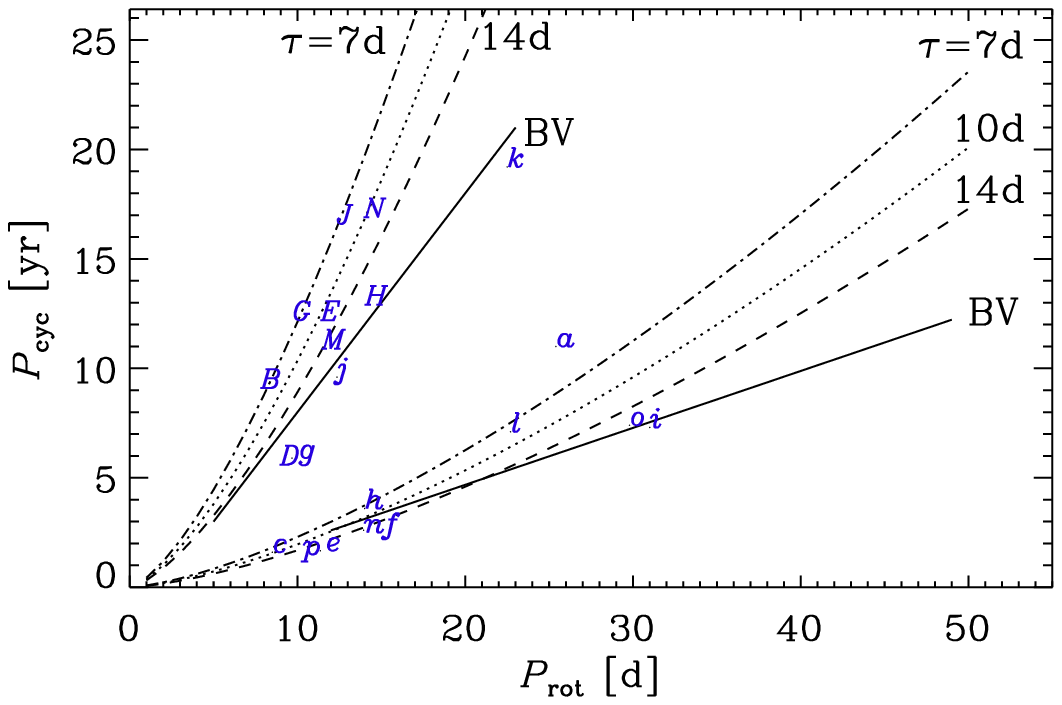}
\end{center}\caption[]{
BV diagram for the F and G~dwarfs listed in \Tab{Gstars}.
Otherwise like \Fig{pKBV}.
}\label{pGBV}\end{figure}

\begin{figure*}[t!]\begin{center}
\includegraphics[width=\textwidth]{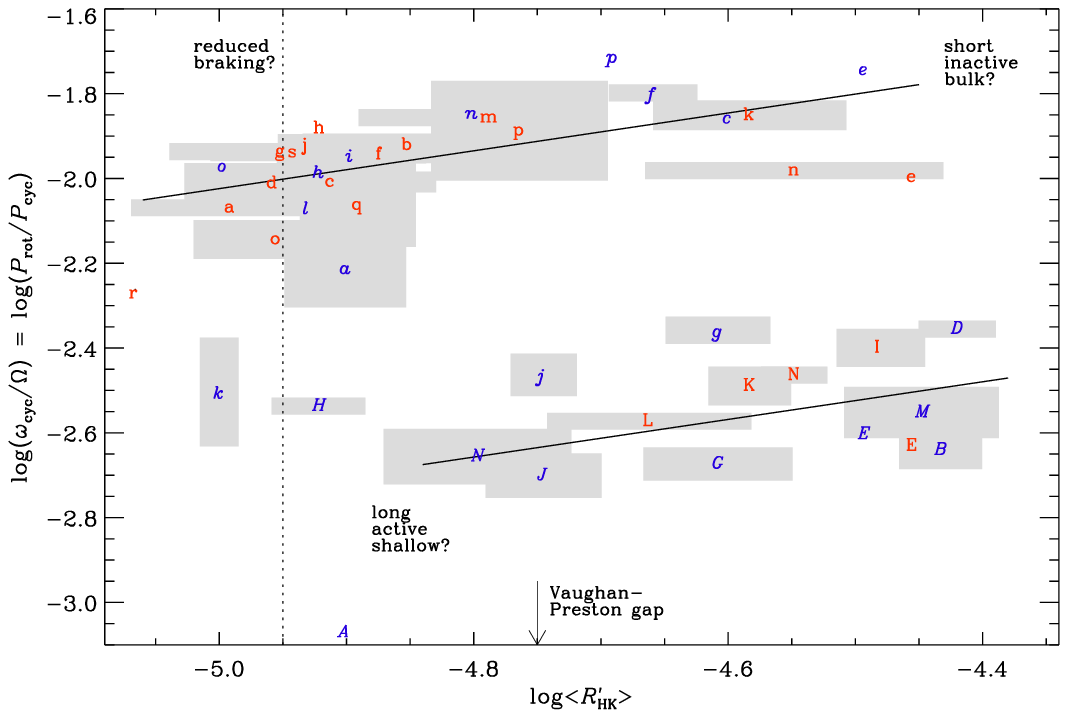}
\end{center}\caption[]{
BST diagram for the F, G, and K~dwarfs showing
$\omega_{\rm cyc}^{\rm I}/\Omega$ (lowercase symbols) and
$\omega_{\rm cyc}^{\rm A}/\Omega$ (uppercase symbols) versus
$\log\bra{R'_{\rm HK}}$ for the K~dwarfs (red roman symbols)
listed in \Tab{Kstars} and F and G~dwarfs (blue italics symbols)
listed in \Tab{Gstars}.
The upper and lower solid lines correspond to fits
to the groups of short and long cycle periods on the
branches of inactive and active stars, respectively.
The gray boxes behind the letters indicate the
statistical error in the period ratio and the spread of
$\log R'_{\rm HK}$ between cycle minimum and maximum.
The vertical dotted line at $\log R'_{\rm HK}=-4.95$
corresponds to $P_{\rm rot}/\tau\approx2.1$, which
\cite{vanSaders2016} identified as the position beyond
which magnetic braking is reduced.
The Vaughan--Preston gap at $\log R'_{\rm HK}=-4.75$ is
marked by an arrow.
}\label{pKG}\end{figure*}

\section{Discussion of individual stars}
\label{Discussion}

We now discuss individual stars, referring to their
red roman symbols for K~dwarfs and blue italics symbols
for F and G~dwarfs in \Fig{pKG}.
We frequently compare the observed cycle periods with
the computed ones.

\begin{table}[b!]\caption{Stars with double periods and their ages.
}\vspace{12pt}\centerline{\begin{tabular}{lll}
F~dwarfs & G~dwarfs & K~dwarfs \\
\hline
(HD~17051) ($0.6\Gyr$)&HD~78366  ($0.8\Gyr$)&HD~22049   ($0.6\Gyr$)\\
HD~114710  ($1.7\Gyr$)&HD~30495  ($1.1\Gyr$)&HD~165341A ($2.0\Gyr$)\\
HD~100180  ($2.3\Gyr$)&(HD~76151) ($1.6\Gyr$)&HD~149661  ($2.1\Gyr$)\\
                      &HD~190406 ($1.8\Gyr$)&                     \\
\label{DoubleCycles}\end{tabular}}
\end{table}

\subsection{Stars with two cycle periods}
\label{StarsWithTwoCyclePeriods}

Looking at \Tabs{Kstars}{Gstars}, there are eight cases of stars
with double periods (three K~dwarfs, four G~dwarfs, and two F~dwarfs).
In addition, we discuss the F~dwarf HD~17051 ($\iota$~Hor)
and the G~dwarf HD~76151,
for which longer cycle periods have tentatively been detected.
The ages of these ten stars are between $0.6$ and $2.3\Gyr$;
see \Tab{DoubleCycles}.
We begin by discussing first the G~dwarfs, where the phenomenon
of two periods is particularly striking.

{\bf HD~78366} (blue $G/g$ symbols) is a young ($0.8\Gyr$) active G0V star
with $\log\bra{R'_{\rm HK}}\approx-4.42$. It has a longer period
of $12.2\yr$ (good), which is close to the computed
value of $9.9\yr$ for the active branch, but the shorter one of $5.9\yr$ (fair)
is much longer than the computed value of $1.9\yr$.
The evidence for the shorter $5.9\yr$ period hinges on two pronounced
``spikes'' in the time series at times when the longer cycle is close
to minimum.
The periodogram of \cite{Egeland17} shows a peak at $5.9\yr$ and also one
at $19.8\yr$.
However, using Zeeman Doppler Imaging, \cite{Morgenthaler11} found two
reversals between 2008 and 2011, which would be compatible with our
computed cycle period of $1.9\yr$.
They also quoted a slightly longer rotation period of $11.4\days$ instead
of $9.7\days$.
This would move the longer cycle period (blue $G$ symbol)
even closer to the A branch.
This star from the Mount Wilson HK project is also being monitored with
the Solar-Stellar Spectrograph\footnote{\url{http://www2.lowell.edu/users/jch/sss/}} (SSS)
at Lowell Observatory.
Looking at \Fig{pG}, there is some similarity to HD~114710 (blue $J/j$ symbols);
in both cases, the shorter period is just one half of
the longer one, so these data points are still close to the active branch.

{\bf HD~30495} (=~58~Eri, blue $E/e$ symbols), is a young ($1.1\Gyr$)
variable G1V star of BY~Draconis type, and is considered a solar analog.
No cycle periods were given by \cite{Bal+95}, and only recently
did \cite{Egeland15} discover a short cycle period of $1.67\yr$ and a
long one of $12.2\yr$.
Both values agree well with the computed ones of $2.0$ and $10.3\yr$,
respectively.
Although the longer cycle period is apparent in the more recent time trace
of \cite{Hall07}, the shorter one is not so clear owing to poor cadence.
With $\log\bra{R'_{\rm HK}}\approx-4.49$, this star is well above the
Vaughan--Preston gap.

{\bf HD~190406} (=~15~Sge, blue $N/n$ symbols) is a solar analog
with an age of $2.3\Gyr$ and two cycle periods.
It is similar to HD~30495, except that it is below the Vaughan--Preston gap
and has $\log\bra{R'_{\rm HK}}\approx-4.80$.
\cite{Bal+95} determined a short cycle of $2.6\yr$ (fair)
and a long one of $16.9\yr$ (good).
Both values agree well with the computed periods of $3.3\yr$
and $17.3\yr$, respectively.
\cite{Egeland17} found $1.6$, $2.6$, and $18.7\yr$ from his periodograms.
The shorter cycle period is also seen in the more recent
observations by \cite{Hall07}.
The short-term Fourier transforms of \cite{Olah_etal16} clearly show
power for both periods, but with different strengths at different times.

{\bf HD~114710} (=~$\beta$~Com, blue $J/j$ symbols) is an F9V star
($1.7\Gyr$) with two cycle periods that were determined by \cite{Bal+95}
to be $9.6\yr$ (fair) and $16.6\yr$ (good).
With $\log\bra{R'_{\rm HK}}\approx-4.75$ it is in the Vaughan--Preston gap.
The computed periods are $2.8\yr$ and $14.5\yr$, respectively.
Although the longer period is close to the computed one, the shorter one
is much longer.
However, looking at the time trace in \cite{Bal+95}, the observational cadence
is insufficient to discern short periods in the $2$--$3\yr$ range.
Therefore, it is conceivable that the shorter period of HD~114710 might not
have been determined correctly.
\cite{Egeland17} did find shorter periods of $5.1$ and $6.5\yr$
in his periodograms.
This star from the Mount Wilson HK project continues
to be monitored with the 
SSS, thereby also allowing better verification also of the
longer cycle period in the future.

{\bf HD~100180} (=~88~Leo, blue $H/h$ symbols) is an inactive
($\log\bra{R'_{\rm HK}}\approx-4.92$) F5V star ($2.3\Gyr$) with two cycle periods
that \cite{Bal+95} determined to be $3.6\yr$ (fair) and $12.6\yr$ (fair).
In this case, the shorter period agrees well with the computed one
of $3.7\yr$, but the longer one turns out to be $19.7\yr$.
However, given that the time series covered only 25 years,
we cannot exclude that the actual period was longer.
This impression is confirmed by direct inspection of the time series;
see \cite{Olah_etal16}, who quote $16.6$--$9.85\yr$ for the range of
the longer cycle period.

{\bf HD~22049} (=~$\epsilon$~Eri, red E/e symbols) is a very young ($0.6\Gyr$)
and active ($\log\bra{R'_{\rm HK}}\approx-4.46$) K2 star, whose
computed cycle periods of $1.8\yr$ and $9.7\yr$ are a bit shorter
than the observed ones of $2.9\yr$ and $12.7\yr$ \citep{Metcalfe13}.

{\bf HD~165341A} (=~70~Oph~A, red N/n symbols) is an active
($\log\bra{R'_{\rm HK}}\approx-4.55$) K1V star with an age of $2.0\Gyr$,
whose longer cycle period of $15.5\yr$ is a bit shorter than the
computed one of $19.1\yr$, whereas the shorter one of $5.1\yr$ is longer
than the computed $3.6\yr$ period.

{\bf HD~149661} (red k/K symbols) is an active
($\log\bra{R'_{\rm HK}}\approx-4.58$) K2V star with an age of $2.1\Gyr$.
The longer of the two periods is $17.4\yr$, which is again a bit
shorter than the computed value of $21\yr$.
We note in passing that \cite{SB99} quoted a cycle period of $16.2\yr$.
The shorter period of $4\yr$ agrees with the computed value.
The monitoring of this star is being continued with the SSS.

{\bf HD~17051} (=~$\iota$~Hor, blue $c$ symbol) is a young ($0.6\Gyr$)
F8V star with one of the shortest cycle periods ever detected
\citep[$1.6\yr$;][]{Metcalfe10}.
This agrees with the computed value.
In \Tab{Gstars} it is listed with a single period (which is why it is
listed in parentheses in \Tab{DoubleCycles}), but recent evidence now suggests
the presence of a longer cycle of $\sim5\yr$ as well \citep{Flores17}.
The computed value is $8.6\yr$.
Strengthening the evidence for the existence of this longer cycle period
should be an important goal for future observations.

For the {\bf Sun} (blue $a$ symbol), we have adopted a cycle period of
$11\yr$ instead of the $10\yr$ that was determined by \cite{Bal+95}
from their limited $\bra{R'_{\rm HK}}$ time series.
As secondary period (blue $A$ symbol),
we took an estimated $80\yr$ Gleissberg cycle period.
The computed value is only about half as long.
However, the Sun is much older than the other stars with double periods,
so the $80\yr$ Gleissberg period may not be a proper secondary cycle
in the same sense.

\subsection{Stars with only a short cycle}

We consider stars whose frequency ratios $\omega_{\rm cyc}/\Omega$ are
close to those on the inactive or short cycle branch.
The $\bra{R'_{\rm HK}}$ values are mostly below the Vaughan--Preston gap
and older than $3\Gyr$.
Exceptions are HD~76151 ($1.6\Gyr$) and KIC~10644253 ($0.9\Gyr$),
with ages for which there are many stars with longer cycles
(\Sec{StarsWithTwoCyclePeriods}).
For HD~76151, \cite{Egeland17} reported a $20.9\yr$ period, which is
close to the computed value of $16.1\yr$, but for KIC~10644253,
the observations are not long enough to confirm a longer period.
In \Tab{DoubleCycles}, HD~76151 has been listed in parentheses.

{\bf HD~201091} (=~61~Cyg~A, red p symbol) is an inactive K5V star ($3.3\Gyr$)
with a $7.3\yr$ cycle period that agrees with the computed value of $8\yr$.
Its surface field geometry has been assessed using Zeeman Doppler Imaging
and is found to be solar-like \citep{BSaikia16}.
There is no evidence for a long-term cycle.

{\bf HD~10476} (=~107~Psc, red c symbol) is a K1V star ($4.9\Gyr$) with a
$9.6\yr$ cycle period -- in agreement with the computed $9.3\yr$ period.
We note in passing that the rotation period used here is $35.2\days$,
as given in BST and \cite{SB99}, but BV listed $38.2\days$, which may
have been a typo.

{\bf HD~81809} (red h symbol) is an old ($6.6\Gyr$) and inactive star
with only one period,
which is a bit shorter ($8.2\yr$) than the computed one of $10.7\yr$.
\cite{Bal+95} classified it as a solar-like G2V star with $B-V=0.64$,
but suggested that the dominant component of this binary might be a
K0V~dwarf, for which BST used $B-V=0.80$.
In fact, with $B-V=0.64$, the value of $\tau$ would be too small and
\Eq{RHK_relation} would result in a value of $\bra{R'_{\rm HK}}$
that is too small, compared to the actual one.
\cite{Egeland17} also quotes $B-V=0.64$ for the dominant component.

{\bf HD~128620} (=~$\alpha$~Cen~A, blue $k$ symbol) is an old ($5.4\Gyr$)
G2V star in the southern hemisphere, and was therefore not included in the
original sample of \cite{Bal+95}.
It has one of the most extensive records in the X-ray \citep{Ayres14}
as well as in the far ultraviolet \citep{Ayr15} wavelengths.
Its cycle period of $19.2\yr$ is slightly closer to the $34\yr$ period computed
for the A~branch than the $6.5\yr$ value for the I~branch.
As will be discussed in \Sec{Metallicity}, its estimated $\bra{R'_{\rm HK}}$
value of $-5.00$ is $0.2$ dex smaller than expected, based on its
rotation period.
Neither circular nor linear polarization has been
detected, indicating the absence of a net longitudinal
magnetic field stronger than $0.2\G$ \citep{Koc11}.
It is fairly metal-rich, with [Fe/H]=0.23.

{\bf HD~128621} (=~$\alpha$~Cen~B, red j symbol) is a K1V star
(also $5.4\Gyr$, but with [Fe/H]=0.27).
It is about 0.1 dex above the I branch in \Fig{pKG} and its cycle period
of $8.1\yr$ is slightly below the computed one of $9.7\yr$.

{\bf HD~201092} (red p symbol) with $B-V=1.37$ is the coolest
K~dwarf ($3.2\Gyr$) in our sample.
Its primary cycle period of $11.7\yr$ is close to
the computed value of $8\yr$.

{\bf HD~219834A} (=~94~Aqr, red r symbol) is the least active cycling star
in our sample ($\log R'_{\rm HK}=-5.07$).
\cite{Bal+95} classified it as K5V dwarf, but \cite{San+10}
determined it to be a G6/G8IV subgiant.
Its age is $6.4\Gyr$ and its cycle period of $21\yr$ exceeds the
computed value of
$13\yr$. At this low activity level, one expects reduced magnetic braking
to have set in \citep{MEvS16}, which could be responsible for
departures from the computed value.
This is expected to occur at a critical value of
$P_{\rm rot}/\tau\approx2.1$ \citep{vanSaders2016}, which, using
\Eq{RHK_relation}, corresponds to $\log R'_{\rm HK}=-4.95$.
This value is marked in \Fig{pKG} by a vertical dotted line.

{\bf HD~103095} (blue $i$ symbol) is an extremely metal-poor G8V star
($4.6\Gyr$) with ${\rm[Fe/H]}=-1.36$.
The cycle period of $7.3\yr$ agrees well with the computed one of
$8.1\yr$.

{\bf HD~76151} (blue $f$ symbol) is an active ($\log R'_{\rm HK}=-4.66$)
G3V star ($1.6\Gyr$) with only a short cycle period of $2.5\yr$.
This agrees with the computed one of $3.1\yr$.
A longer period of $16.1\yr$ is computed, which is close to the
$20.9\yr$ period found by \cite{Egeland17}, but it was not reported
by \cite{Bal+95}.
On the other hand, \cite{Egeland17} reported a $5.0\yr$ period, but his
periodogram also shows a peak at $2.6\yr$.

{\bf HD~146233} (=~18~Sco, blue $l$ symbol) is the currently accepted
best bright solar twin ($4.1\Gyr$).
Its cycle period of $7.1\yr$ \citep{Hall07b}
agrees well with the computed value of $6.1\yr$.
\cite{Egeland17} found $1.9$, $6.5$, and $13.9\yr$ periods.

{\bf HD~166620} (red o symbol) is a slowly rotating inactive K5~dwarf ($6.2\Gyr$)
with a cycle period of $15.8\yr$, which is below the computed $11.7\yr$ value.
It has $\log\bra{R'_{\rm HK}}=-4.96$, which is slightly (0.1~dex) below
the value expected for its slow rotation period of $42.4\days$.
In all other aspects, it is unexceptional; it has a similar age
and $\bra{R'_{\rm HK}}$ value to HD~16160 (red d symbol),
which has an even longer rotation period of $48\days$.

{\bf KIC~8006161} (=~HD~173701, blue $o$ symbol), is a G8V star ($4.6\Gyr$)
with a short period of $7.7\yr$, which is somewhat shorter than the computed
$8.6\yr$.
With [Fe/H]~=~0.34, this star has the largest metallicity in our sample.

{\bf KIC~10644253} (blue $p$ symbol) is a young ($0.9\Gyr$) solar analog of
spectral type G5V \citep{Salabert16a}, with a $1.5\yr$ cycle period
\citep{Salabert16b} determined from seismology,
which is shorter than the computed period of $2.3\yr$.
The computed value for the longer cycle of $12.1\yr$ is not observed,
but looking for a longer cycle should be a priority for
future observations.

There are several other K~dwarfs with single short cycles that are
not discussed here in detail: HD~3651, HD~4628, HD~16160, HD~26965,
HD~32147, HD~160346, HD~201091, and HD~219834B.
In all of these cases, the observed cycle periods are close to the
computed values.
Their ages are in the range $3.3$--$7.2\Gyr$.

\subsection{Stars with only a long cycle}

Stars with only a long cycle period lie on the A branch
and are in that sense active stars.
The stars HD~1835, HD~152391, and HD~20630 are well above the Vaughan--Preston
gap with $\log\bra{R'_{\rm HK}}$ in a narrow range between $-4.45$ and $-4.42$
and have ages from $0.5$ to $0.8\Gyr$.
They all have computed cycle periods on the short cycle branch between
$0.8$ and $1.4\yr$ that would be interesting to look for in future
observing programs.

{\bf HD~1835} (blue $b$ symbol) is a young G3V star ($0.5\Gyr$) with a cycle period of
$9.1\yr$, which is longer than the computed value of $6.4\yr$ for the active
branch.
\cite{Egeland17} found a slightly shorter value of $7.8\yr$
as well as a longer period of $20.8\yr$, which is not, however, computed.
Instead, there is a shorter computed cycle period of $0.8\yr$
for the inactive branch, which is not observed.

{\bf HD~152391} (blue $m$ symbol) is an active G7V star ($0.8\Gyr$) with one
period of $10.9\yr$, which falls on the active branch with a computed
value of $9.6\yr$.
The shorter period of $1.4\yr$ is not observed.
This star from the Mount Wilson HK project is also being
monitored with the SSS.

{\bf HD~20630} (blue $d$ symbol) is a G5V star ($0.7\Gyr$) with a $5.7\yr$
cycle, which is close to the computed value of $7.5\yr$ for the active
branch.
The computed shorter cycle period of $1.0\yr$ for the inactive branch
is not observed.
\cite{Egeland17} found $5.7$, $14.2$, and $35.8\yr$ periods.

Other stars with only a single long cycle include the K~dwarfs
{\bf HD~115404} and {\bf HD~156026}, with ages $1.4$ and $1.3\Gyr$,
respectively.
As can be seen from the 36 year time series analyzed by \cite{Olah_etal16},
a longer time series would be desirable in both cases.
Of these stars, however, only HD~115404 continues to be monitored with
the SSS.

\subsection{Stars without cycles}

{\bf HD~186408} (=~16~Cyg~A, blue $q$ symbol) and
{\bf HD~186427} (=~16~Cyg~B, blue $r$ symbol) are Sun-like dwarfs,
but at an age of $7\Gyr$ they have no cycle.
\Tab{Gstars} gives computed cycle periods of $7.6$ and $7.3\yr$
that are meaningless in this case.
As we shall discuss later, their $\bra{R'_{\rm HK}}$ values are
smaller than what is expected based on their $\tau/P_{\rm rot}$
values, which is a clear indication that they have experienced
reduced magnetic braking \citep{MvS17}.

\section{Dependencies of the residuals on other quantities}
\label{Dependence_RHK}

\Eqs{oO_RHK_relation}{RHK_relation} define two distinct residuals:
\EQ
\Delta_i=\log(\omega_{\rm cyc}/\Omega)
-\log\left(b_i\bra{R'_{\rm HK}}^{\nu_i}\right)
\label{Deltai}
\EN
with $i={\rm I}$ and $i={\rm A}$ for the inactive and active branches,
respectively, and
\EQ
c=\log\bra{R'_{\rm HK}} - \log (\tau/P_{\rm rot})
\label{cResidual}
\EN
for both branches.
Both $\Delta_i$ and $c$ should be constants, i.e., they should
not depend systematically on any quantity unless there are indeed
additional dependences; for example, on the depth of the convection
zone, the metallicity, or the age.
In the following, we consider these possibilities.

\subsection{Dependence on convection zone thickness}

\cite{TRB88} suspected that $\omega_{\rm cyc}/\Omega$ depends
on the depth of the convection zone.
This possibility arises because the $\alpha$ effect is proportional
to the correlation length $\ell$ of the turbulence with
$\alpha\approx\Omega\ell$ times some quenching function, as discussed
in the introduction.
By parameterizing the differential rotation in terms of the local
double-logarithmic derivative, $q\equiv\dd\ln\Omega/\dd\ln r$, we have
$\Omega'=q\Omega/r$, where $r$ is the radius, so \Eq{omcyc}
yields $\omega_{\rm cyc}/\Omega\approx|q\ell/r|^{1/2}$.
In the near-surface shear layer of the Sun, we have $q\approx-1$
\citep{BSG14}; however, in deeper layers, it is positive and of appreciable
amplitude only in the tachocline \citep{Schou98}.
If the turbulence governing the $\alpha$ effect is characterized
by giant cells of a size comparable to the depth $d$ of the convection zone,
one would expect $\omega_{\rm cyc}/\Omega \propto (d/R)^{1/2}$;
see \Eq{omcyc}.
Evidence from stellar cycle data was presented in \cite{TRB88},
but they ignored the additional dependence on $\bra{R'_{\rm HK}}$.
Therefore, one should consider the dependence of the residual $\Delta_i$
on $d/R$.

\begin{figure}[t!]\begin{center}
\includegraphics[width=\columnwidth]{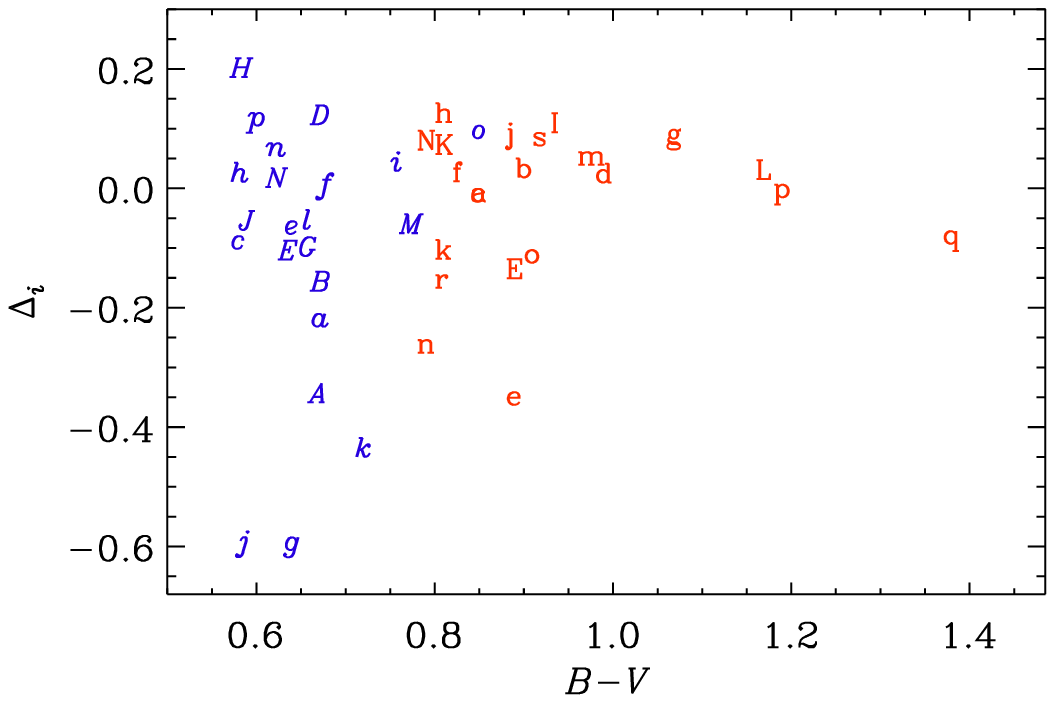}
\end{center}\caption[]{
The residual $\Delta_i$ versus $B-V$ for the K~dwarfs (red symbols)
listed in \Tab{Kstars}, and the F and G~dwarfs (blue symbols) listed
in \Tab{Gstars}.
Note the increase of scatter for small values of $B-V$.
}\label{pKG_vs_BV_comp}\end{figure}

As a very preliminary measure of $d/R$, let us first look at the dependence
of $\Delta_i$ on $B-V$; see \Fig{pKG_vs_BV_comp}.
It is immediately evident that the scatter is much larger for
F and G~dwarfs on the left.
For $B-V>0.9$, the scatter is significantly smaller.
This justifies our approach of considering the K~dwarfs separately
at first instance.
Other than that, no systematic dependence on $B-V$ is found.

\begin{figure}[t!]\begin{center}
\includegraphics[width=\columnwidth]{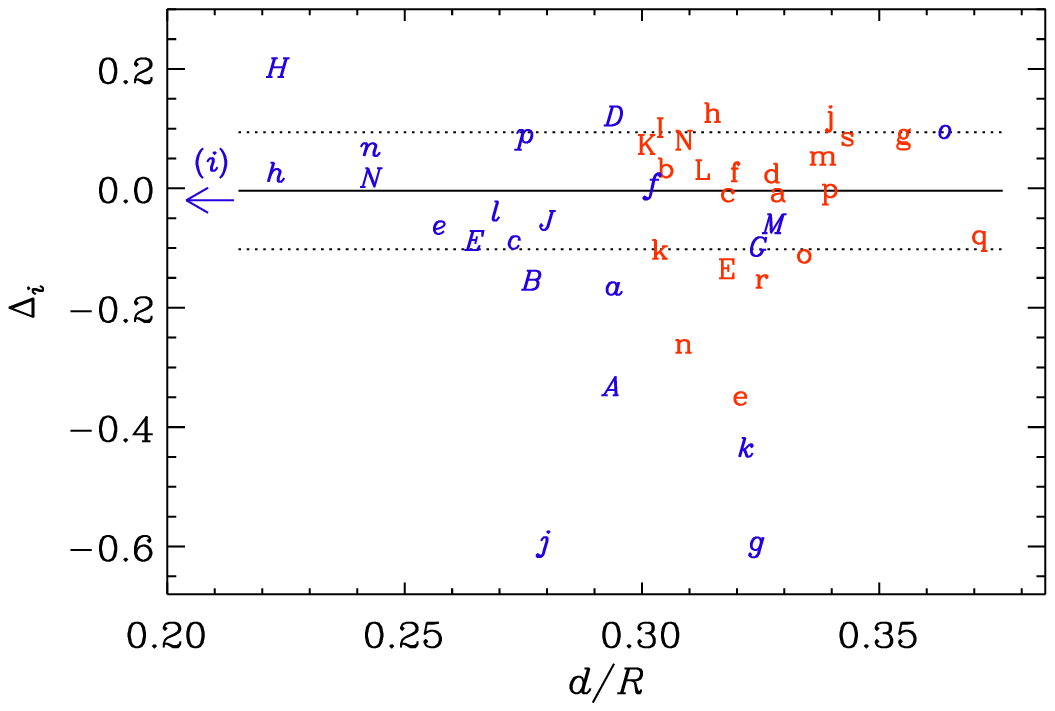}
\end{center}\caption[]{
The residual $\Delta_i$ versus $d/R$ for the K~dwarfs (red symbols)
listed in \Tab{Kstars}, and the F and G~dwarfs (blue symbols) listed
in \Tab{Gstars}.
The solid horizontal line marks the average, and the dotted horizontal
lines the standard deviation.
As elsewhere, the red e and n symbols, as well as the blue $A$, $H$, $E$,
$j$, and $g$ symbols are excluded from the fit. The blue $k$ symbol is,
however, included.
}\label{pKG_vs_dR_comp}\end{figure}

Next, we compute the depth of the convection zone for all of the stars
with a metallicity measurement.
From $B-V$, we compute the mass using the \citet{Noyes84}
relation for main-sequence stars. We then look for the closest model in terms of 
mass and metallicity in the grid of models computed by 
\citet{vSP12}, giving an estimate of the depth of the convection zone as
listed in \Tabs{Kstars}{Gstars}.
The relation of the residuals with $d/R$ is shown in \Fig{pKG_vs_dR_comp} 
for the K~dwarfs of \Tab{Kstars}
in red roman symbols, together with the F and G~dwarfs of \Tab{Gstars}
in blue italics symbols.

It turns out that, within error bars, $\Delta_i$ does not vary
systematically with $d/R$.
Thus, if $\omega_{\rm cyc}/\Omega$ is indeed independent of $d/R$,
we can expect that the correlation length of the dynamo is the same
in stars with different values of $d/R$.
This is remarkable, and suggests that changes in the depth of the convection
zone are not important for the operation of stellar dynamos.
However, the values of $\omega_{\rm cyc}/\Omega$ are different by a
factor $\tilde{b}_{\rm I}/\tilde{b}_{\rm A}\approx6$ on the two branches.
This could be related to separate types of dynamos in a star, which are
characterized by different correlation lengths.
On the branch with a cycle 6 times longer, $(\omega_{\rm cyc}/\Omega)^2$
is 36 times smaller, and thus the correlation length is expected to
be 36 times shorter.
Shorter correlation lengths are normally associated with stellar surface
layers, where the pressure scale height is smaller.
This conclusion agrees with that of BV, who associated the
active branch with a dynamo operating in the near-surface shear layer.
This may well be compatible with the interpretation of \cite{See16},
who found that the magnetic topology on the active branch has a strong
toroidal component.
Strong toroidal fields are suggestive of a  shallow origin, whereas
deeply rooted dynamos are expected to exhibit mostly poloidal fields
at the surface.

The idea of different dynamos was already proposed by \cite{DMR81}
to explain the Vaughan--Preston gap.
Although \cite{DMR81} argued in favor of a change of magnetic field topology,
BV talked explicitly about the simultaneous operation of two dynamos.
She imagined both of them being interface dynamos, but that the dynamo
on the inactive branch would be affected by mixing in deeper layers,
whereas that on the active branch would operate preferentially in the
near-surface shear layers of rapidly rotating G~dwarfs.
Meanwhile, global convective dynamo simulations
\citep{BBBMT10,Racine,KMB12} have produced magnetic fields
in the bulk of the convection zone.
Some of the simulations have demonstrated the simultaneous occurrence
of multiple dynamo periods within the same star \citep{Kapy16,BSCC16}.
In the simulations of \cite{KMCWB13}, I and A branches have been found
that were separated by a factor of four in $\omega_{\rm cyc}/\Omega$.
These simulations produced longer periods near the bottom of
the convection zone.
By contrast, earlier mean-field models of \cite{CTM00,CTVM01} resulted
in shorter periods, which would be in agreement with our interpretation.
It would therefore be interesting to see whether any of these models can
produce cycle diagnostics similar to those discussed in the present paper.

\subsection{Dependence on metallicity}
\label{Metallicity}

\begin{figure}[t!]\begin{center}
\includegraphics[width=\columnwidth]{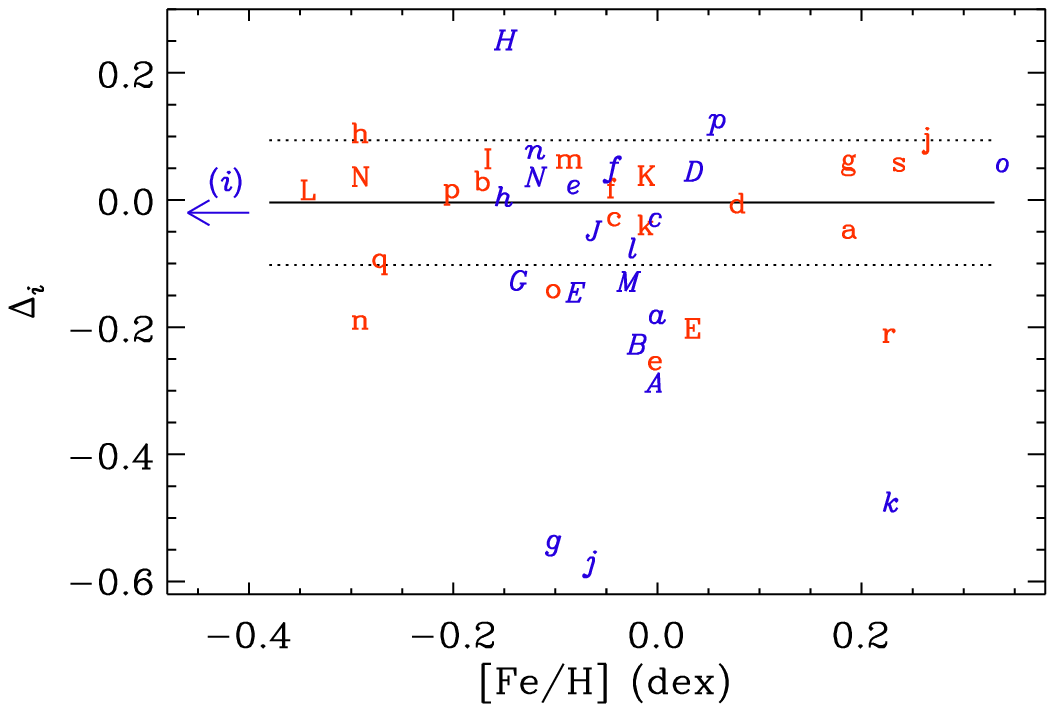}
\end{center}\caption[]{
The residual $\Delta_i$ versus [Fe/H].
Otherwise like \Fig{pKG_vs_dR_comp}.
\vspace{12pt}}\label{pKGmetal_comp}\end{figure}

To see whether unusual metallicities can be responsible for some of the
systematic departures between the observed and computed cycle periods,
we plot the dependence of $\Delta_i$ on [Fe/H] in \Fig{pKGmetal_comp}.
It turns out that the stars with the largest departures from $\Delta_i=0$
all have moderate values of [Fe/H]; see HD~78366 (blue $g$ symbol) and HD~114710
(blue $j$ symbol), for which $\Delta_i\approx-0.55\dex$ and HD~100180
(blue $H$ symbol), for which $\Delta_i\approx+0.25\dex$.
Thus, we conclude that there is no systematic trend between $\Delta_i$
and [Fe/H].
Conversely, the star with the smallest metallicity (HD~103095,
blue $i$ symbol) has ${\rm[Fe/H]}=-1.36$, but the cycle period of $7.3\yr$
agrees well with the computed one of $8.2\yr$.

\begin{figure}[t!]\begin{center}
\includegraphics[width=\columnwidth]{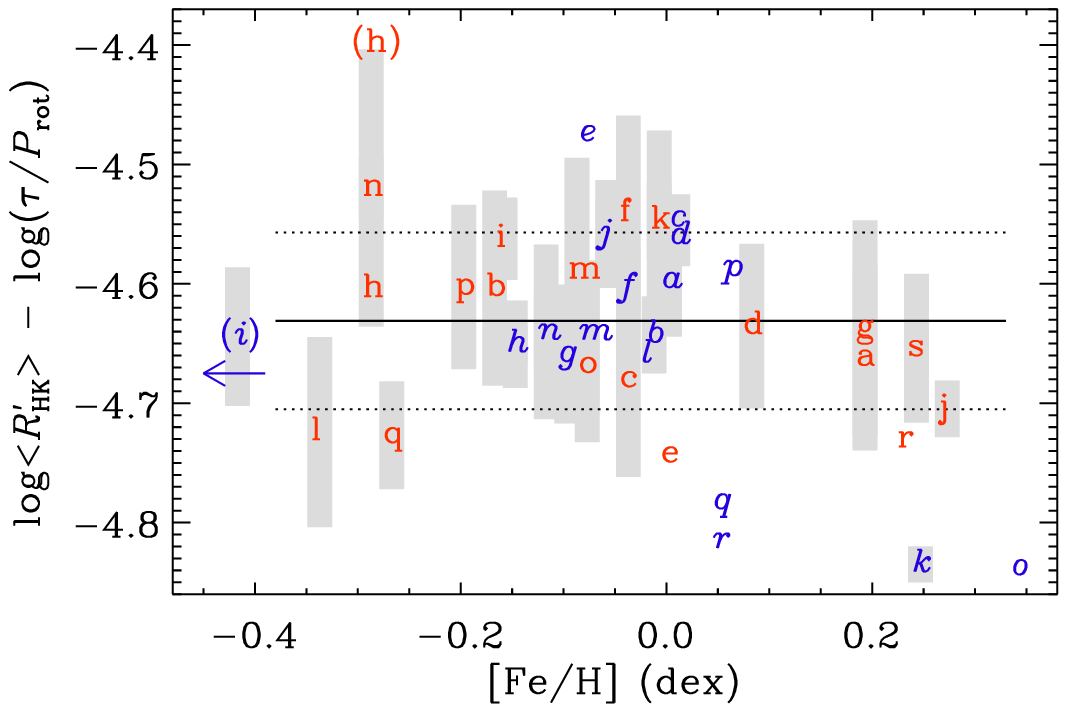}
\end{center}\caption[]{
Plot of the residual $\log\bra{R'_{\rm HK}}-\log(\tau/P_{\rm rot})$
versus [Fe/H]
for the K~dwarfs of \Tab{Kstars} in red roman symbols, together
with the F and G~dwarfs of \Tab{Gstars} in blue italics symbols.
The cycle amplitude is indicated by the vertical extent of the gray boxes.
The solid horizontal line marks the average, and the dotted horizontal
lines the standard deviation.
The red h symbol marks the location of HD~81809 for $B-V=0.80$
whereas ``(h)'' indicates where the location would be if the value
$B-V=0.64$ from \cite{Egeland17} were adopted.
\vspace{12pt}}\label{pRHK_residual_FeH}\end{figure}

Stars with higher (lower) metallicity have a larger (smaller) convection
zone thickness, thereby mimicking properties of stars of later (earlier)
spectral type that have larger (smaller) values of $\tau$. This could
imply larger (smaller) values of $\bra{R'_{\rm HK}}$, and hence
smaller (larger) values of $\Delta_i$.
One would then expect a negative slope in \Fig{pKGmetal_comp}.
No reliable slope is found, although it is interesting to note that
this would be in the right direction to explain the departure found
for HD~128620 (blue $k$ symbol).
This argument would assume that the approximately linear relation between
$\bra{R'_{\rm HK}}$ and $\tau/P_{\rm rot}$ remains valid and that the
$c$ in \Eq{RHK_relation} does not itself depend on [Fe/H].
\Fig{pRHK_residual_FeH} shows that this is indeed the case.

In this connection, we recall that, for HD~81809 (red h symbol), we assumed
$B-V=0.80$, which might not be justified \citep{Egeland17}.
If we were to use $B-V=0.64$, the value of $\tau$ would be smaller,
and therefore $c$ would be larger.
This point is marked in \Fig{pRHK_residual_FeH} with ``(h),'' which
exceeds the average value of $c$ by more than $0.2\dex$.
On the other hand, if HD~81809 is indeed a subgiant, as argued by
\cite{Egeland17}, then \Eq{RHK_relation} might not be valid without
additional adjustments.

\subsection{Dependence on age}

\begin{figure}[t!]\begin{center}
\includegraphics[width=\columnwidth]{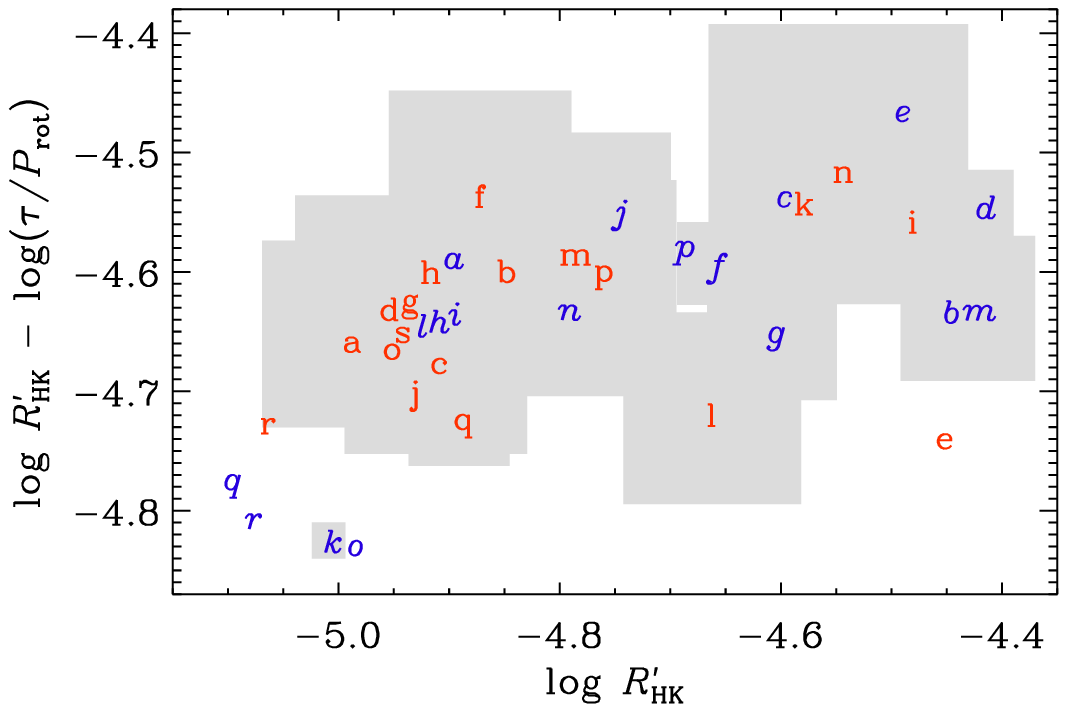}
\end{center}\caption[]{
Residual $\log\bra{R'_{\rm HK}}-\log(\tau/P_{\rm rot})$
versus $\log\bra{R'_{\rm HK}}$
for the K~dwarfs in red roman characters, together
with the F and G~dwarfs in blue italics characters.
The cycle amplitude is indicated by the vertical extent of the gray boxes,
whereas the horizontal extent denotes the spread of $\log\bra{R'_{\rm HK}}$
between cycle minimum and maximum.
Note the systematic departures for small values of $\log\bra{R'_{\rm HK}}$.
}\label{pRHK_residual_RHK}\end{figure}

Stellar activity decreases monotonically with age for dwarfs, so
it is instructive to plot the residual $c$ from \Eq{cResidual}
versus $\log\bra{R'_{\rm HK}}$.
Figure~\ref{pRHK_residual_RHK} shows that \Eq{RHK_relation} begins
to fail for the smallest values of $\log\bra{R'_{\rm HK}}<-5.0$.
Stars with very low activity are ``superrotating,'' i.e., they
rotate faster than expected based on their activity level.
This is again a clear indication that these stars, notably
16~Cyg~A and B (blue $q$ and $r$ symbols),
$\alpha$~Cen A (blue $k$ symbol), and
KIC~8006161 (blue $o$ symbol), have experienced reduced magnetic braking
as a consequence of the large-scale cyclic dynamo having started
to shut down \citep{KN17}.
For 16~Cyg~A and B (blue $q$ and $r$ symbols in \Fig{pRHK_residual_RHK}),
this has already happened.

As discussed by BV, the coexistence of long and short period cycles in
some stars suggests that multiple stellar dynamos can operate
simultaneously.
As we have seen from \Tab{DoubleCycles}, this is a possibility for all
stars younger than about $2.3\Gyr$.
This interpretation offers a fresh opportunity
to consider the evolution of stellar cycles in the BST diagram as
stars move from high levels of activity (right side of \Fig{pKG}) toward
lower activity states (left side of \Fig{pKG}). 

Evidently, long period
cycles are dominant for stars on the active side of
the Vaughan--Preston gap, whereas short period cycles dominate
on the inactive side. This idea may help to explain
some of the outliers discussed above. In particular, the short
period cycle in HD~22049 (red e symbol), which is very young ($0.6\Gyr$),
may be operating outside the optimal range of
the underlying dynamo, so it falls below the pattern established by
other cycles on the inactive branch.
By contrast, HD~165341A (red n symbol) and HD~149661 (red k symbol) are
older, and their short cycles agree reasonably well with the computed ones.
HD~30495 (blue $e$ symbol) is also fairly young ($1.1\Gyr$), but its short
cycle period agrees with stars on the inactive branch.
Evolving toward the Vaughan--Preston gap, short period cycles begin to appear 
slightly above the inactive branch (blue $f$ and $p$ symbols). At comparable 
activity levels, we see coexisting long and short cycles for HD~78366 (blue 
$G/g$ symbols) and HD~114710 (blue $J/j$ symbols). The long cycles in these 
stars fall on the active branch, whereas the short cycles appear to be 
outliers.
Just across the Vaughan--Preston gap,
we find HD~190406 (blue $N/n$ symbols), which is the only star in the sample 
with long and short period cycles that both fall directly onto their
respective branches.
At an age of $2.1\Gyr$, it is no longer very young.

Moving to lower activity levels we find HD~100180 (blue $H/h$ symbols)
at $2.3\Gyr$,
which shows a short period cycle on the inactive branch and a long
period cycle that falls above the active
branch. This could signal the operation of the underlying dynamo beyond
its optimal range, although the long period cycles might simply exceed
the reach of current time domain surveys. Below a critical activity
level, even the inactive branch may no longer sustain coherent stellar
cycles. Based on the positions of two solar analogs in the BV diagram,
\cite{MCvS17} suggest that the short period cycle in HD~146233 (blue $l$
symbol, $4.1\Gyr$) may evolve away from the inactive branch toward the longer
period cycle in HD~128620 (blue $k$ symbol, $5.4\Gyr$) before the cycle disappears
entirely, as in the old solar analogs 16~Cyg~A and B \citep{Hall07,MCD15}
at $7\Gyr$.
This transition appears to coincide with the reduced magnetic braking
suggested by \cite{vanSaders2016}, possibly due to a reconfiguration of
the field toward smaller spatial scales as the global dynamo begins to
shut down \citep{MEvS16}. The Sun (blue $a$ symbol, $4.6\Gyr$) appears to be
on the threshold of this transition, particularly during solar minimum. The
absence of additional points at lower activity levels may simply reflect
the disappearance of cycles below this threshold \citep{MvS17}. Future observations of
inactive {\it Kepler} stars will help to clarify this picture.

\begin{figure}[t!]\begin{center}
\includegraphics[width=\columnwidth]{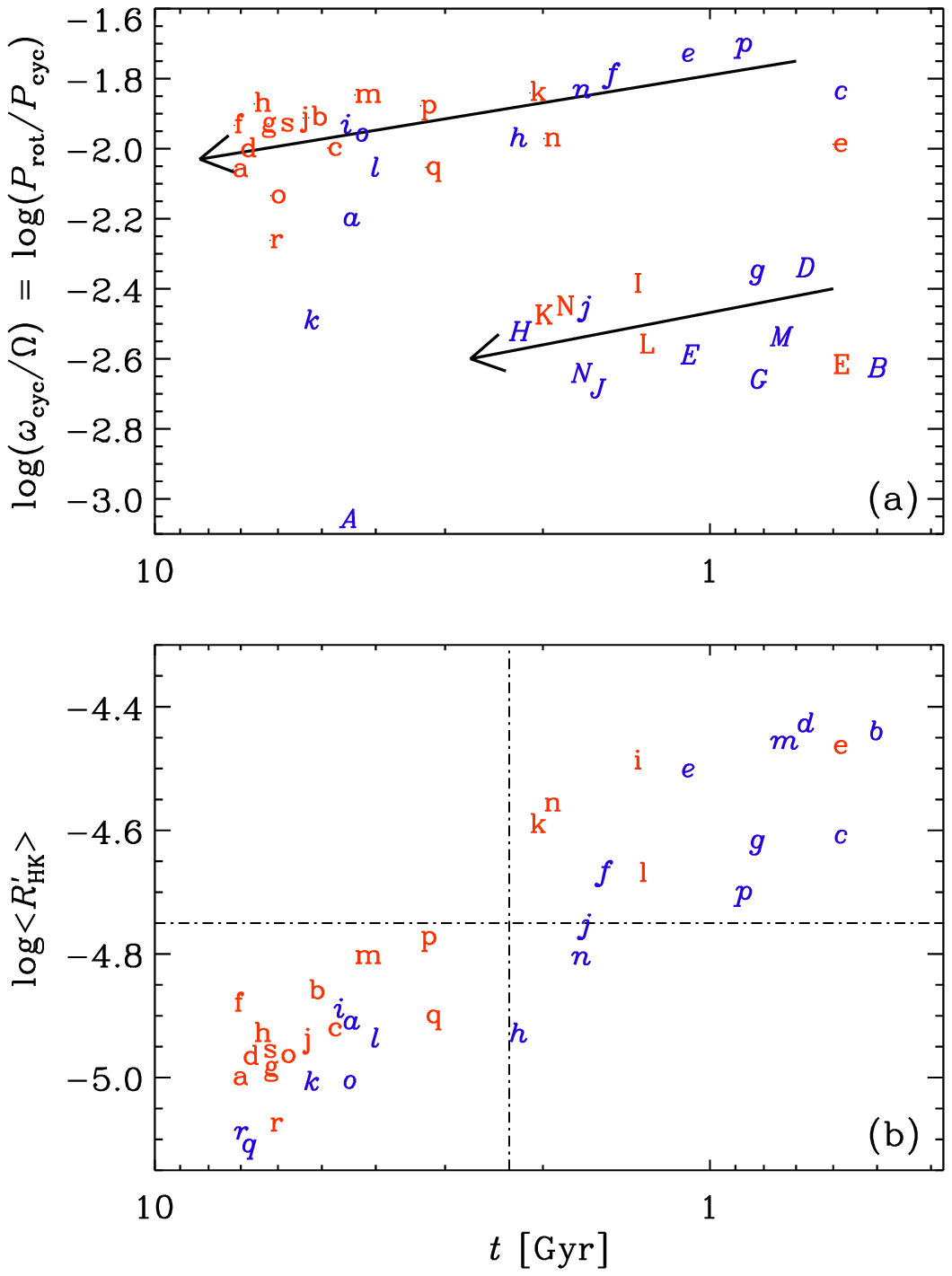}
\end{center}\caption[]{
(a) Frequency ratio versus age.
Note that age increases to the left, to facilitate comparison
with the BST diagram.
The arrows indicate tentative evolutionary tracks.
(b) $\bra{R'_{\rm HK}}$ versus age.
}\label{pKG_vs_age}\end{figure}

BST assumed that $\bra{R'_{\rm HK}}$ was a reasonable proxy of age.
Gyrochronology and asteroseismology allow us now to express the
frequency ratio directly in terms of age; see \Sec{StellarAges}.
A BST plot with age on the abscissa is shown in \Figp{pKG_vs_age}{a}.
Compared to the BST plot in \Fig{pKG}, the scatter is larger.
However, both plots agree, in that the long and short cycle branches
overlap in the ranges
\EQ
-4.85\leq\bra{R'_{\rm HK}}\leq-4.45,\quad
0.6\leq t/{\rm Gyr}\leq2.3,
\EN
where both branches are populated.
This is different from the original BST plot, where the overlap
is almost absent.

In \Figp{pKG_vs_age}{b}, we show $\bra{R'_{\rm HK}}$ versus age.
It shows that all stars older than $2.3\Gyr$ have an
$\bra{R'_{\rm HK}}$ value below the Vaughan--Preston gap.
Most of the younger stars are above the Vaughan--Preston gap,
except HD~100180 (blue $h$ symbol), HD~114710 (blue $j$ symbol),
and HD~190406 (blue $n$ symbol), which are below the gap.
These exceptions are all G~dwarfs with two cycle periods.
This shows that {\em all} stars with long cycle periods are younger
than $2.3\Gyr$ -- even the rather inactive stars
HD~190406 (blue $n$ symbol) and HD~100180 (blue $h$ symbol)
in the lower right quadrant of \Fig{pKG_vs_age}{b}.

The reverse is not true: many stars with short cycle periods
can still be young and may only have a short cycle, namely
HD~17051 (blue $c$ symbol with $P_{\rm cyc}=1.6\yr$).
HD~76151 (blue $f$ symbol with $P_{\rm cyc}=2.5\yr$), and
KIC~10644253 (blue $p$ symbol with $P_{\rm cyc}=1.5\yr$).
However, it would be important to keep looking for longer secondary
cycle periods, which may already have been found in the case of
HD~76151 \citep{Egeland17}.

\cite{Pace_etal09} have suggested that the evolution across the
Vaughan--Preston gap may be rather abrupt.
Our \Figp{pKG_vs_age}{b} could be compatible with this idea,
especially for the F~dwarfs, where we find several stars in a narrow
interval around $1.7\Gyr$ outlining a jump in $\bra{R'_{\rm HK}}$.
For the K~dwarfs, this may also happen, but it would be somewhat
later---between $2$ and $3\Gyr$, although in our sample has no
K~dwarfs in that age range.

\section{Conclusions}

The present work supports the idea that longer and shorter stellar cycles
tend to fall on one of two universal branches in the BST diagram.
In the corresponding BV diagram, these lines are not universal, but
correspond to a family of lines for different values of the turnover time.
For K~dwarfs, the values of $\tau$ lie within a relatively narrow range
between $17$ and $23\days$, so the statistical quality between the BST
and BV diagrams is similar.
For F and G~stars, on the other hand, the scatter is significantly
larger in both the BST and BV diagrams.

It was already known that the BST diagram can be cast into an
evolutionary diagram.
However, unlike the previous interpretation whereby young stars
would evolve along the active or long cycle period branch, we now
see that {\em all} stars younger than $2.3\Gyr$ are capable of
exhibiting longer and shorter cycle periods.
This implies that the solar Gleissberg cycle would not be a
secondary cycle in the same sense, because the Sun is older than $2.3\Gyr$.

Our work has allowed us to compute secondary cycle
periods that could be longer or shorter than the observed ones.
It will be interesting to see whether this is borne out by
future observations.
For some stars, the possibility of as yet undetected shorter cycle
periods in the $1$--$2\yr$ range is now a possibility, notably for the
G dwarfs HD~1835, HD~20630, and HD~152391.
Less clear is the situation for the K~dwarfs HD~115404 and HD~156026,
for which shorter cycle periods in the $3$--$4\yr$ range are possible.
On the other hand, for the G dwarfs HD~76151 and KIC~10644253,
for which shorter cycle periods have been detected, longer periods
in the $12$--$16\yr$ range are possible, and may have been already found
in the case of HD~76151 \citep{Egeland17}.
Similarly, for $\iota$~Hor (HD~17051), the star with a
short cycle of $1.6\Gyr$, a longer $\sim5\yr$ cycle may have already emerged
\citep{Flores17}.
If this possibility gets confirmed,
it will be interesting to see how the measured period would
compare with the computed cycle of $8.5\yr$.
Such a time frame would be more manageable than those associated
with the secondary Gleissberg-type cycles of solar-like stars.
On the other hand, our work now suggests that the Gleissberg
cycle of the Sun is distinct from the longer secondary cycles discussed
here for our sample of ten stars.

We have found no systematic dependence of the period ratio or
the activity level on metallicity or $d/R$.
This suggests that these aspects of the dynamo are not strongly
affected by the depth of the star's convection zone.
This would be more suggestive of distributed dynamos, although it
would be premature to make strong claims whereas the solar dynamo
is not yet well understood.
For the most inactive stars, cyclic dynamo activity has
ceased and the chromospheric activity level has dropped below the value
that is expected based on the star's rotation rate.

\acknowledgements
We thank the anonymous referee for useful comments and
Ricky Egeland for interesting discussions.
This work has been supported in part by
the NSF Astronomy and Astrophysics Grants Program (grant 1615100),
the Research Council of Norway under the FRINATEK (grant 231444),
the Swedish Research Council (grant 621-2011-5076)
and by NASA grant NNX15AF13G.

%r e f

\end{document}